# Finite Element Simulations of an Elasto-Viscoplastic Model for Clay

**Mohammad N. Islam [1,*], Carthigesu T. Gnanendran [1] and Mehrdad Massoudi [2]**

[1] School of Engineering and Information Technology, University of New South Wales, Canberra, ACT 2612, Australia; r.gnanendran@adfa.edu.au

[2] US Department of Energy, National Energy Technology Laboratory (NETL), P.O. BOX 10940, 626 Cochrans Mill Road, Pittsburgh, PA 15236, USA; Mehrdad.Massoudi@netl.doe.gov

* Correspondence: nislamce@gmail.com

**Abstract:** In this paper, we develop an elasto-viscoplastic (EVP) model for clay using the non-associated flow rule. This is accomplished by using a modified form of the Perzyna's overstressed EVP theory, the critical state soil mechanics, and the multi-surface theory. The new model includes six parameters, five of which are identical to those in the critical state soil mechanics model. The other parameter is the generalized nonlinear secondary compression index. The EVP model was implemented in a nonlinear coupled consolidated code using a finite-element numerical algorithm (AFENA). We then tested the model for different clays, such as the Osaka clay, the San Francisco Bay Mud clay, the Kaolin clay, and the Hong Kong Marine Deposit clay. The numerical results show good agreement with the experimental data.

## 1. Introduction

Saito and Uezawa Saito and Uezawa [1] showed that the Takabayama landslide in Japan was mainly due to the time dependent deformation of the clay, resulting in the failure of the slope. In 1963, the failure of the Vayont reservoir in Italy, cost more than 3000 lives [2]; this was again due primarily to the creep of clay. These and many other examples in geotechnical engineering, reservoir geomechanics, wellbore plugging in petroleum and mining industry, etc., have made studying and modeling of clay a very important topic.

Clay is considered to be a multi-component system composed of solid particles, and fluid components (gas and/or liquid) [3]. For many contaminated soils, there might be additional components, such as bubbles, oil spill, etc. In a partially saturated clay, the void space is filled with gas and/or liquid, and, in a fully saturated clay, only a liquid phase is present. Removal of any fluid from this system depends on the magnitude of the external load and the drainage path, the imposed duration, and the properties of the clay. When an external load is applied, pressure is developed in the fluid, which may dissipate instantaneously or gradually [4]. On the other hand, for a solid medium with non-crushable fine materials, when the fluid pressure dissipation is negligible, the settlement of clay does not end. In this case, creep may continue for a long time under a constant pressure [5]. This process constitutes a complex hydro-mechanical phenomenon [6,7].

Many constitutive models have been developed for clay. These range from simple elastic models to very complicated elasto-viscoplastic models. An overview can be found in the works of Owen and Hinton [8] and Desai and Siriwardane [9]. Chaboche [10] also presented a review of constitutive theories for plastic and viscoplastic type models (see also Tatsuoka et al. [11]). Liingaard et al. [12] suggested that the time-dependent viscous type models can be separated into: (i) empirical models (Singh and Mitchell [13]); (ii) rheological models (Feda [14]); and (iii) models considering "stress-strain-time" (Adachi and Okano [15]). The first two categories are important for the basic understanding of the behavior of clay. In clay-based research and its finite element implementation, the third category is most popular. For the remainder of this paper, the discussion is limited to the elasto-viscoplastic (EVP) type models. In the formulation of the EVP models, the strain rate consists of an elastic component and an inelastic component. To obtain an expression for the latter, it is essential to include the time-dependent viscous property in the model formulation. Sekiguchi [16] subdivided EVP models into: (a) the overstressed type [15]; (b) the non-stationary flow surface type (Sekiguchi [17]); and (c) others, such as the Bounding surface model [18] and Borja model [19]. Details of EVP models and their strength, as well as their limitations, can be found in the work by Liingaard et al. [12].

In this paper, we develop a new EVP model with a non-associated flow rule, considering Perzyna's viscoplastic theory [20] and the MCC framework [21]. The time-dependent viscous aspect is incorporated using the Borja and Kavazanjian [19] approach. The new model requires six parameters. The current study is a generalization of the previous work by Islam and Gnanendran [22], who showed the strengths and the limitations of the existing methods when using the two-surface approach in the EVP models by incorporating the associated flow rule. Here, we use the non-associated flow rule and the three-surface approach. We also look at the Osaka clay, the San Francisco Bay Mud clay, the Kaolin clay, and the Hong Kong Marine Deposit clay in a triaxial loading situation. The wide range of experiments included the drained and the undrained tests, the strain rate test, the relaxation test, and the over consolidation effect tests. The model was implemented in a finite-element code using the numerical algorithm (AFENA) [23]. The pertinent details of the finite element implementation and the development of the non-associated flow rule are discussed in this paper.

## 2. Basic Equations

We consider clay to be a mixture of two constituents composed of solid particles and a liquid. We use the basic ideas and principles of continuum theories of mixtures, where electromagnetic and thermo-chemical effects are ignored (see Atkin and Craine [24] and Bowen [25]). The motion for each component is given through a mapping

$$\boldsymbol{x} = \boldsymbol{\chi}^i(\boldsymbol{X}^i, t) \quad i = s, f \tag{1}$$

where $\boldsymbol{\chi}^i$ designates the mapping, and $i$ refers to $s$ (solid component-clay) and $f$ (fluid component). For clay, the kinematical quantities of interests are the displacement vector $\boldsymbol{u}$, the velocity vector $\boldsymbol{v}^s$, the deformation gradient tensor $\boldsymbol{F}$, and the linearized strain $\boldsymbol{\varepsilon}^s$:

$$\boldsymbol{u}(\boldsymbol{x}, t) = \boldsymbol{x}(\boldsymbol{x}, t) - \boldsymbol{X}^s \tag{2}$$

$$\boldsymbol{v}^s = \frac{\partial \boldsymbol{X}^s}{\partial t} \tag{3}$$

$$\boldsymbol{F} = \frac{\partial \boldsymbol{x}^s}{\partial \boldsymbol{X}^s} \tag{4}$$

$$\boldsymbol{\varepsilon}^s = \frac{1}{2}\left[\left(\frac{\partial \boldsymbol{u}}{\partial \boldsymbol{x}}\right) + \left(\frac{\partial \boldsymbol{u}}{\partial \boldsymbol{x}}\right)^T\right] \tag{5}$$

where superscript $T$ designates the transpose of a tensor. If $V_0$ is the volume of the particles before the deformation and $V$ is the volume after the deformation, then a quantity of interest is the volumetric strain, also called the dilatation

$$\varepsilon_v = tr\,\boldsymbol{\varepsilon} = \varepsilon_{kk} = \frac{V - V_0}{V_0} \tag{6}$$

Furthermore, in elastoplastic theories, for small deformations, we use $\dot{\boldsymbol{\varepsilon}}$ instead of the symmetric part of the velocity gradient $\boldsymbol{D}^s$. That is, for small deformation, we assume $\dot{\boldsymbol{\varepsilon}} = \boldsymbol{D}^s$, where

$$\dot{\varepsilon}_{ij} = \frac{1}{2}\left(\frac{\partial \dot{u}_i}{\partial x_j} + \frac{\partial \dot{u}_j}{\partial x_i}\right) \tag{7}$$

For the fluid component, the kinematical quantities of interest are the velocity vector $\boldsymbol{v}^f$, the gradient of the velocity $\boldsymbol{L}$, and its symmetric part $\boldsymbol{D}$, which are given by, respectively,

$$\boldsymbol{v}^f = \frac{\partial \boldsymbol{X}^f}{\partial t} \tag{8}$$

$$\boldsymbol{L} = \frac{\partial \boldsymbol{v}^f}{\partial \boldsymbol{x}} \tag{9}$$

$$\boldsymbol{D} = \frac{1}{2}[\boldsymbol{L} + \boldsymbol{L}^T] \tag{10}$$

We assume that the density of the solid and fluid components in their current configuration are $\rho_s$ and $\rho_f$, respectively, while, in their pure (reference) state, i.e., before mixing, they are given as $\rho_R^s$and $\rho_R^f$. The density and the velocity of the mixture are then defined as

$$\rho(\boldsymbol{x},t) = \rho_s(\boldsymbol{x},t) + \rho_f(\boldsymbol{x},t) \tag{11}$$

$$\boldsymbol{v}(\boldsymbol{x},t) = \frac{1}{\rho}\left[\rho_s \boldsymbol{v}^s + \rho_f \boldsymbol{v}^f\right] \tag{12}$$

Note that Equation (12) is one of the many possibilities for defining a mixture velocity. We can also define a relative velocity $\boldsymbol{w}$ such that

$$\boldsymbol{w}(\boldsymbol{x},t) = \boldsymbol{v}^f(\boldsymbol{x},t) - \boldsymbol{v}^s(\boldsymbol{x},t) \tag{13}$$

The densities in the current and reference configuration are related through the kinematical field $\varphi(\boldsymbol{x},t)$, called the porosity, such that

$$\rho_s = (1-\varphi)\rho_R^s \tag{14}$$

$$\rho_f = \varphi\rho_R^f \tag{15}$$

where $0 \le \varphi(\boldsymbol{x},t) \le \varphi_{max} < 1$. Thus, when $\varphi = 1$, we have a fluid with no pores and no particles. Note that this automatically assumes that the mixture is saturated. Otherwise, the porosities are constrained by $\varphi_s + \varphi_f \le 1$. In geomechanics-related problems, we sometimes use the void ratio $e$, which is related to the porosity $\varphi$, through

$$e = \frac{\varphi}{1-\varphi} \tag{16}$$

Assuming no inter-conversion of mass between clay and the liquid in the pores, the equations for the balance of mass for the two constituents are given by

$$\frac{\partial \rho^i}{\partial t} + div\,(\rho^i \boldsymbol{v}^i) = 0; \qquad i = s,f \tag{17}$$

where $\frac{\partial(.)}{\partial t}$ is the time derivative and $div$ is the divergence operator. The balance of linear momentum for the two components are given by

$$\rho^i \frac{d^i \boldsymbol{v}^i}{dt} = div\,\boldsymbol{\sigma}^i + \rho^i \boldsymbol{b}^i + \boldsymbol{\mathcal{M}}^i\,; \qquad i = s,f \tag{18}$$

where $\boldsymbol{b}^i$ is the external body force, $\boldsymbol{\mathcal{M}}^i$ is the interaction forces, $\boldsymbol{\sigma}^i$ designates the partial stress tensor and $\frac{d^i(.)}{dt} = \frac{\partial(.)}{\partial t} + [grad\,(.)].\,\boldsymbol{v}^i$ is the total time derivative. We can define the stress tensor for the mixture (the total stress) as

$$\boldsymbol{\sigma} = \boldsymbol{\sigma}^s + \boldsymbol{\sigma}^f \tag{19}$$

If $\boldsymbol{t}^s$ denotes the traction vector of the solid component on the boundary, then

$$\boldsymbol{t}^s = \boldsymbol{\sigma}^s \boldsymbol{n}^s \tag{20}$$

where $\boldsymbol{n}^s$ is the unit vector acting on the boundary. In accordance with the basic principles of mixture theory, advocated by Truesdell [26], if we add the two equations for the conservation of mass or the conservation of the linear momentum, we obtain the balance of mass and the linear momentum for the mixture (note that, in this case, $\boldsymbol{\mathcal{M}}^i$ drops out of the equations), due to application of the Newton's third law

$$\boldsymbol{\mathcal{M}}^s = -\boldsymbol{\mathcal{M}}^f \tag{21}$$

Finally, the balance of angular momentum indicates that, in the absence of angular momentum supply, the total stress is symmetric. That is [see Rajagopal [27]]

$$\boldsymbol{\sigma} = \boldsymbol{\sigma}^T \tag{22}$$

Since we consider a non-isothermal problem, we do not discuss the balance of energy or the Second Law of thermodynamics.

## 3. Constitutive Modeling

In this paper, the emphasis is on the modeling of the stress tensor of the solid particles (clay). However, for the closure of the governing equations, we also need constitutive relations for the stress tensor for the fluid, $\boldsymbol{\sigma}^f$, and the interaction forces $\boldsymbol{\mathcal{M}}^i$.

### *3.1. Fluid Component and the Interaction Forces*

The following assumptions, as elaborated in Martins-Costa et al. [28] and Rajagopal [27], would lead to the classical Darcy's equation. The interaction forces designated by $\boldsymbol{\mathcal{M}}^i$, within the context of mixture theory and many of the multiphase theories, are usually based on generalizing the interaction force for very special cases, such as the Stokes drag on a single spherical particle.

We assume that the frictional (viscous) forces within the fluid can be ignored and as a result the partial stress tensor for the fluid can be given by a Eulerian fluid model:

$$\boldsymbol{\sigma}^f = -p^f(\rho_f)\boldsymbol{I} \tag{23}$$

where $p^f$ is, in general, a function of density, and $\boldsymbol{I}$ is the identity tensor [note that compressive stresses are assumed to be negative in these theories, whereas in, geomechanics-related problems, the opposite convention is used]. We further assume, as is customary in geomechanics problems and basic flows through porous structures (see Oka and Kimoto [29], p. 34):

$$\boldsymbol{\sigma}^f = -\varphi p\boldsymbol{I} \tag{24}$$

The interaction force is assumed to be

$$\boldsymbol{\mathcal{M}}^f = \alpha(\boldsymbol{v}^f - \boldsymbol{v}^s) \tag{25}$$

where $\alpha$ is a coefficient that can depend on porosity, viscosity, permeability, etc. This is basically a generalization of the Stokes' drag on a single spherical particle. In general, the interaction forces could depend on other kinematical quantities such as the relative acceleration, velocity gradients, etc. (see Massoudi [30,31] for a review of this topic). To obtain the Darcy's equation, we ignore the inertial effects, i.e., we ignore the left-hand side of Equation (18) (when written for the fluid component), and, by using Equations (13) and (23), we obtain the Darcy's equation:

$$-\nabla p^f + \alpha\boldsymbol{w} + \rho_f\boldsymbol{b}^f = 0 \tag{26}$$

where $\rho_f$, as mentioned before, is the density of the fluid. Furthermore, by assuming (see Martins-Costa, et al. [28], Williams [32])

$$\alpha = -\frac{\mu}{\boldsymbol{k}} \tag{27}$$

where $\boldsymbol{k}$ represents the specific permeability, which for anisotropic materials is generally a second-order tensor. Equation (26) can be re-written as

$$\boldsymbol{w} = -\frac{\boldsymbol{k}}{\mu}\left(\nabla p^f - \rho_f\boldsymbol{b}^f\right) \tag{28}$$

In soil mechanics literature, Darcy's equation is sometimes expressed using the concept of hydraulic conductivity ($\boldsymbol{K}$) ( see Bear and Bachmat (1990, p.294)), defined by

$$\boldsymbol{K} = \frac{\boldsymbol{k}\rho_f \boldsymbol{g}}{\mu} \tag{29}$$

Here, $\rho_f \boldsymbol{g}$ represents the volumetric weight of the fluid, $\gamma_f$, which is assumed to act in the vertical direction ($\boldsymbol{b}_f = [0, \gamma_f, 0]^T$). Equation (28) is then re-written as:

$$\boldsymbol{w} = -\frac{\boldsymbol{K}}{\gamma_f}\left(\nabla p^f - \rho_f \boldsymbol{g}\right) \tag{30}$$

We use this form of the equation in our finite element simulation.

### 3.2. Solid Component

The partial stress for the solid component can be defined as (Lewis and Schrefler [33]):

$$\boldsymbol{\sigma}^s = \boldsymbol{\sigma}' - (1 - \varphi)p_f \boldsymbol{I} \tag{31}$$

where $\boldsymbol{\sigma}'$ is the effective stress tensor and $p_f$ is the pore (fluid) pressure. We can relate the total stress tensor of the mixture $\boldsymbol{\sigma}$ [see Equation (19)] to the effective stress tensor $\boldsymbol{\sigma}'$ [by adding Equations (19), (24) and (31)], namely

$$\boldsymbol{\sigma} = \boldsymbol{\sigma}' - p_f \boldsymbol{I} \tag{32}$$

We assume that the strain in clay can be decomposed into an elastic part and a viscoplastic part. In plasticity theory, we find it more convenient to assume this decomposition applies to for the strain rates [see Davis and Selvadurai [34], p.97]

$$\dot{\boldsymbol{\varepsilon}} = \dot{\boldsymbol{\varepsilon}}^e + \dot{\boldsymbol{\varepsilon}}^{vp} \tag{33}$$

We also assume that the elastic part of the strain can be represented by the "small-strain" or the linearized theory of elasticity, where, as customary in soil mechanics, the strain is assumed to depend on the effective stress (Terzaghi [4], pp.11–15), Schofield and Wroth ([35], p. 9). For an isotropic material, using the index notation, the elastic strain is given by (Matsuoka and Sun [36],p. 37)

$$\dot{\varepsilon}_{ij}^e = \frac{1+\nu}{E}\dot{\sigma}_{ij}' - \frac{\nu}{E}(\dot{\sigma}_{kk}')\delta_{ij} \tag{34}$$

Where in accordance with the critical state theory, we assume $E = \frac{3(1-2\nu)(1+e_0)p}{\kappa}$ is the (modified form of the) Young's modulus, $\nu$ is the Poisson's ratio, $\kappa$ is the slope of the unloading and reloading path (see [Matsuoka and Sun [36] (p.35, Figure 2.8); Desai and Sriwardane [9](p. 289, Figure 11.7)), $p = \frac{\sigma_{kk}}{3}$ is the initial mean pressure, $e_0$ is the initial void ratio, and $\delta_{ij}$ is the Kronecker delta ($\delta_{ij} = 1(i = j), 0(i \neq j)$). In a more compact form (using the index notation), Equation (34) can be written as

$$\dot{\varepsilon}_{ij}^e = S_{ijkl}\dot{\sigma}_{kl}' \tag{35}$$

where $S_{ijkl}$ is the fourth-order compliance tensor, related to $C_{ijkl}$ the stiffness tensor. Since we have assumed that the material is isotropic, in short hand notation [recalling Hooke's law $\dot{\sigma}_{kl}' = C_{ijkl}\dot{\varepsilon}_{kl}^e$],

$$S_{ijkl} = \begin{cases} \dfrac{1}{E} & i = j \quad i,j = 1,2,3 \\ \dfrac{\nu}{E} & i \neq j \quad i,j = 1,2,3 \\ G & i = j \quad i,j = 4,5,6 \end{cases} \tag{36}$$

In Equation (35) $\dot{\sigma}_{kl}'$ can be generalized to the case of an elasto-viscoplastic case (see Desai and Sriwardane [9], p. 294, Equation 11.32), i.e.,

$$\dot{\sigma}'_{kl} = D_{ijkl}\left(\dot{\varepsilon}_{ij} - \dot{\varepsilon}_{ij}^{vp}\right) \tag{37}$$

where $D_{ijkl}$ is a fourth-order tensor similar to the elastic moduli.

For the viscoplastic modeling of the strain rate, we start with Perzyna, who used the associated flow rule and assumed the plastic potential function coincides with the loading function [20]. To apply this theory to geomaterials, the main challenge is to define the static loading function [12]. By definition, $f_s$, represents the stress state where the strain rate is assumed to be zero. Here, we assume that the viscoplastic part of the strain rate in Equation (33) is based on Perzyna's approach, where

$$\dot{\varepsilon}_{ij}^{vp} = \langle\psi(F)\rangle \frac{\partial f_p}{\partial \sigma'_{ij}} \quad (38)$$

$$\langle\psi(F)\rangle = \begin{cases} \psi(F) & : F > 0 \\ 0 & : F \leq 0 \end{cases}$$

$$F = \frac{f_l - f_r}{f_r}$$

In the above equations, $\psi$ is the rate sensitivity function, and its functional form can be obtained either experimentally or theoretically; $\langle\ \rangle$ is the Macaulay's bracket (in Equation (38), the Macaulay's bracket ensures that the function inside the bracket only has a value when it is positive, otherwise its magnitude is zero); $F$ is the over-stress function; and $f_p$ is a new term, representing the new potential surface (surfaces of the proposed EVP model are obtained by extending the Modified Cam Clay surface [Roscoe and Burland [21]; in our EVP model, we require a total of three surfaces (see Figure 1): the loading surface $f_l$, the reference surface $f_r$, and the potential surface $f_p$) given by

Potential surface:

$$f_p = p_p^2 - p_{cp}p_p + \left(\frac{q_p}{M}\right)^2 = 0 \quad (39)$$

and $f_l$ and $f_r$ are given by the same expression as those in the Perzyna model. They are the dynamic loading function (the potential surface) and the static loading functions, respectively, given by

Reference surface:

$$f_r = p_r^2 - p_{cr}p_r + \left(\frac{q_r}{M}\right)^2 = 0 \quad (40)$$

Loading surface:

$$f_l = p_l^2 - p_{cl}p_l + \left(\frac{q_l}{M}\right)^2 = 0 \quad (41)$$

where $p = \sigma_{oct} = \frac{\sigma_{kk}}{3}$, $q = \frac{3}{\sqrt{2}}\tau_{oct} = \left[\frac{3}{2}(\sigma_d)_{ij}(\sigma_d)_{ij}\right]^{\frac{1}{2}}$. Similar expressions for $p$ and $q$ (also known as the deviatoric stress) can also be found in the work by Borja and Kavazanjian [19]. In the principal stress space, $\sigma_{oct}$ and $\tau_{oct}$ are the octahedral normal stress and the octahedral shear stress, respectively (see Matsuoka and Sun [36], pp. 29–30). In the above equations, the suffixes $r, l,$ and $p$ represent the reference surface, the loading surface, and the potential surface, respectively. The meaning of these surfaces is shown in Figure 1. At any given time, the reference stress state and the reference surface are known from the laboratory test. The current stress state and the potential stress state are related to the reference stress state through the radial mapping rule, which was proposed by Phillips and Sierakowski [37]. In this paper, we used two image parameters and the details are given in Appendix A. It is worth mentioning that Islam and Gnanendran [22] demonstrated the strengths and the limitations of the existing methods using the two-surface approach in the EVP models. They used the associated flow rule for their EVP model, while, in the present paper, we use the non-associated flow rule and the three-surface approach. In the previous paper, the surface shapes are two ellipses, while, in the present study, we assume all the surfaces are given by a single ellipse, which is close to the MCC surface. To formulate the new EVP model, we assume that the "projection center" is in the origin of the stress space, which is identical to the MCC model. However, to define the surfaces, the expression of the slope of the critical state line $(M)$ is changed with respect to the $b$-value $\left(= \frac{\sigma_2 - \sigma_3}{\sigma_1 - \sigma_3}\right)$ [38] as

$$M = \frac{6\,sin\phi\sqrt{b^2 - b + 1}}{3 + (2b - 1)sin\phi} \tag{42}$$

where $\phi$ is the internal angle of friction at the failure for each $b$-value test, ranging from 0 (triaxial compression) to 1 (triaxial extension).

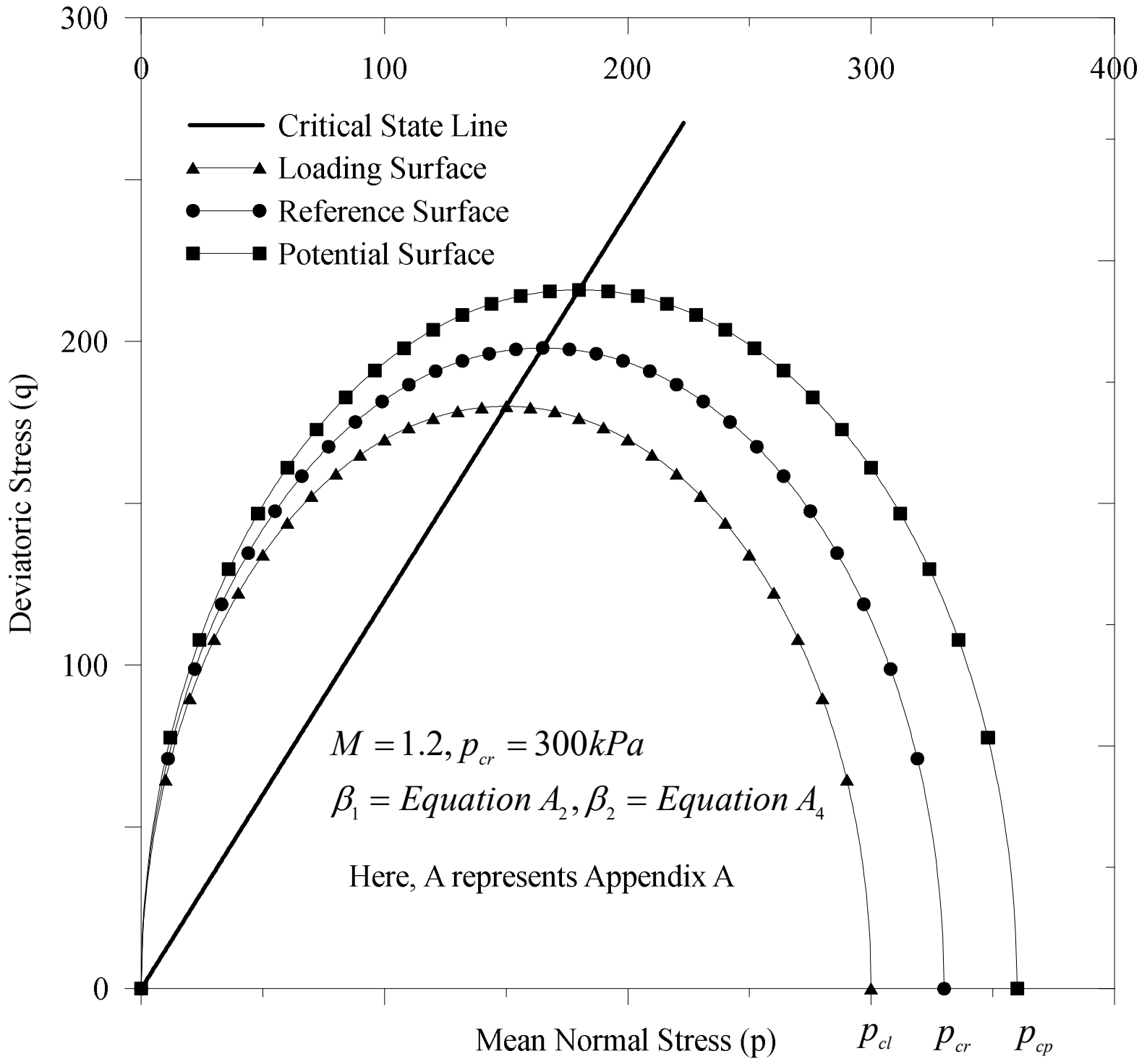

**Figure 1.** Illustration of the reference surface, the loading surface, and the potential surface in the $p - q$ plane.

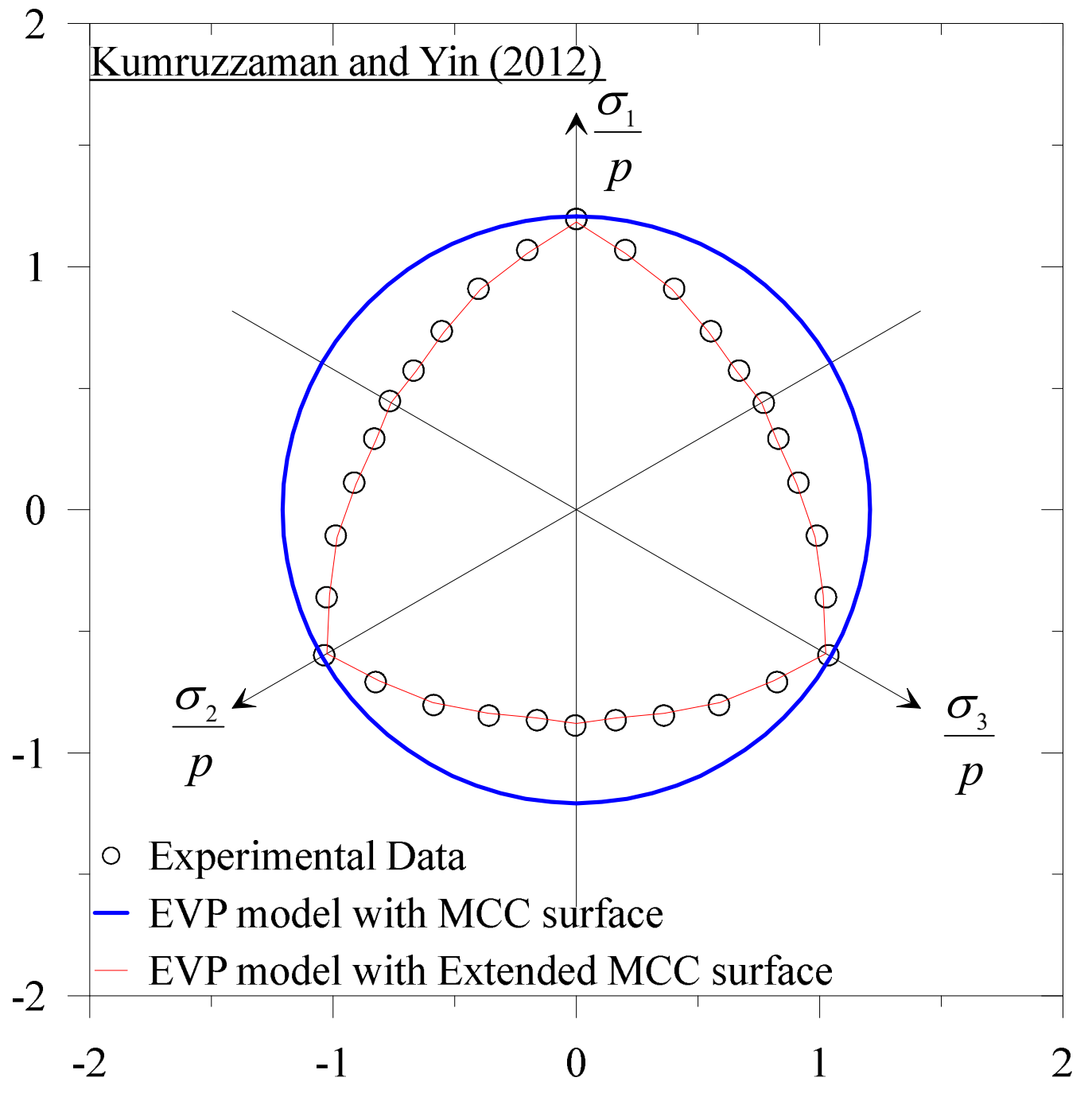

**Figure 2.** Comparison of the MCC surface and the extended surface in the $\pi$-plane.

We have introduced a new parameter $M$ in order to obtain a more realistic surface shape in the $\pi$-plane (see Islam and Gnanendran [22]). It is observed that, in any stress state $(0 < b - value \leq 1)$ other than the triaxial compression state $(b - value = 0)$, the MCC equivalent surface overestimates the stress. For the sake of completeness, in Figure 2, we compare the EVP model based on the MCC surface with the new modified surface presented in this paper. It is observed that the newly extended MCC surface captures and compares well with the experimental results.

In Figure 3, we can see that, at any arbitrary reference time $\bar{t}$, the soil state is at "A", where the corresponding void ratio is $\bar{e}$ .With time changing from $\bar{t}$ to t, due to creep, the soil moves from "A" to "B" where the corresponding void ratio is $e$. Then, the following expressions are obtained from Borja and Kavazanjian [19]:

$$\bar{e} = e_N - \lambda \ln p_{cl} + \kappa \ln\left(\frac{p_{cl}}{p}\right) \tag{43}$$

$$e = e_N - \lambda \ln p_{cr} + \kappa \ln\left(\frac{p_{cr}}{p}\right) \tag{44}$$

where $\lambda$ and $\kappa$ are the slope of the normal consolidation line (the $\lambda$-line) and the unloading-reloading line (the $\kappa$-line), respectively, and $e_N$ is the void ratio corresponding to the $\lambda$-line when $p = 1$ kPa at $\bar{t}$.

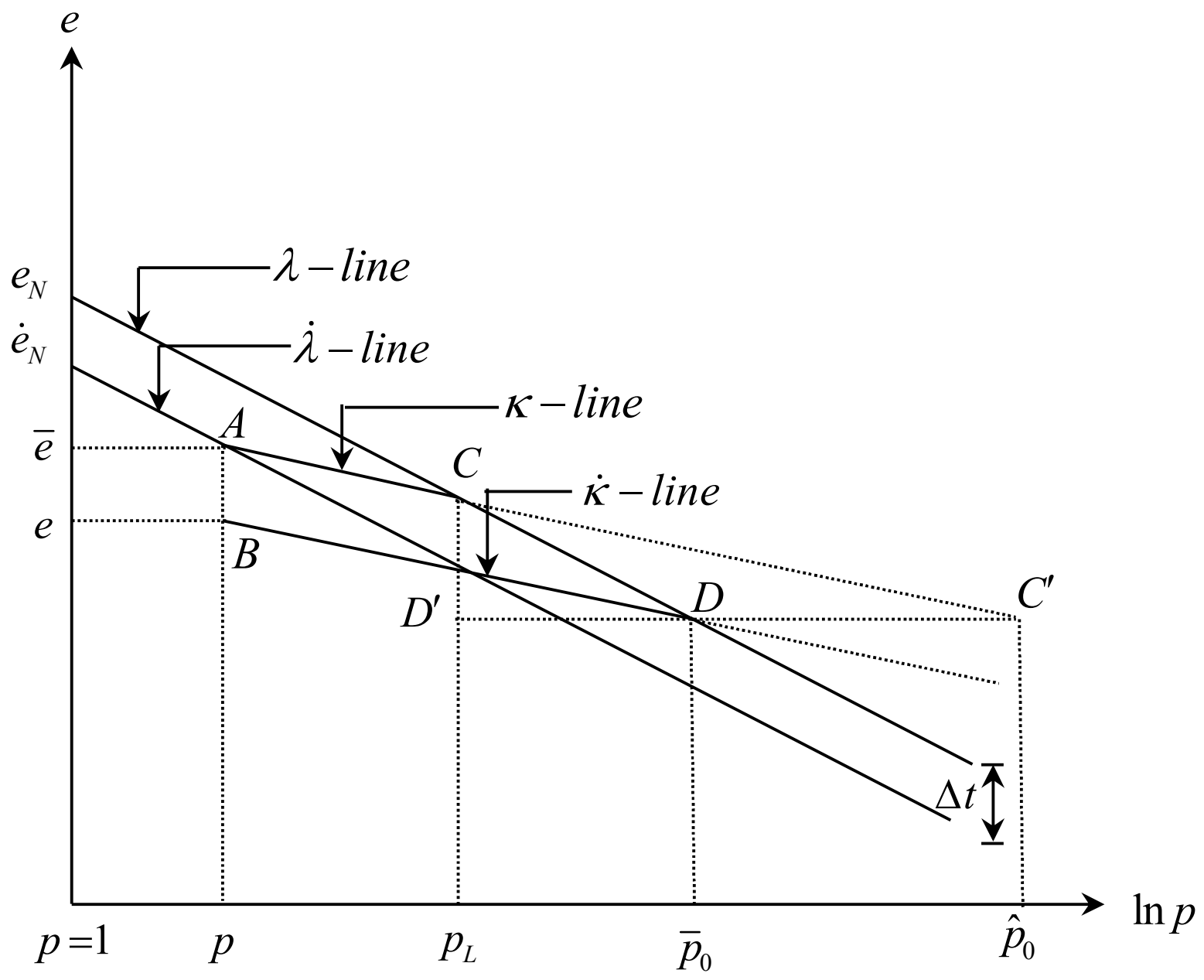

**Figure 3.** Relative locations of $p_{cl}, p_{cr}$ and $p_{cp}$.

It should be mentioned that $\lambda$, $\kappa$ and $e_N$ are the necessary parameters in the EVP model; these can be obtained either from the oedometer test or the triaxial test. The meaning of these parameters is the same as those in the MCC model. As time increases, the $\lambda$-line changes to the $\dot{\lambda}$-line and $e_N$ is transformed to $\dot{e}_N$. The $\dot{\lambda}$-line will generate the new bounding surface. The $\lambda$-line and the $\dot{\lambda}$-line are parallel, as shown by Bjerrum theory [39]. Using Borja and Kavazanjian's [19] concept and the multisurface theory, $\bar{t}$ is the arbitrary time, representing the state of stress prior to the surface evolving. In Equations (40) and (41), $p_{cl}$ and $p_{cr}$ are also known as the creep exclusive preconsolidation pressure and the creep inclusive preconsolidation pressure, respectively (see Islam and Gnanendran [22]). The expression for $p_{cl}$ is similar to the one used in the MCC model:

$$p_{cl} = p + \frac{q^2}{pM^2} \tag{45}$$

After rearranging Equation (44), the expression for $p_{cr}$ can be obtained

$$p_{cr} = exp\left(\frac{e_N - e - \kappa\, lnp}{\lambda - \kappa}\right) \tag{46}$$

From Figure 3 for the definitions of $\lambda$ and $\kappa$, we can obtain an expression for $p_{cp}$:

$$p_{cp} = \frac{p_{cr}^{\frac{\lambda}{\kappa}}}{p_{cl}^{\frac{\lambda-\kappa}{\kappa}}} \tag{47}$$

The detailed derivation of $\langle\psi(F)\rangle$ and $\dot{\varepsilon}_{ij}^{vp}$ are presented in Appendices B and C, respectively. In closing this section, we need to mention that the overstressed EVP models are usually based on the Perzyna's theory [20] in combination with the critical state soil mechanics theory, e.g., the Modified Cam Clay (MCC) model [21]. In these approaches, the viscous nature of the EVP model is introduced in the theory using a secondary compression index, a creep function and a relaxation function. However, in most cases, to reduce the complexity of the model and to minimize the number of parameters, the EVP models are developed considering the associated flow rule, where it is assumed that the yield surface is identical to the potential surface. To capture the behavior of geomaterials, using the non-associated flow rule is essential [40]. Depending on the application of the Critical States Soil Model (CSSM) in the EVP models using the non-associated flow rule, there are two approaches we can consider: (i) those with critical state [40]; and (ii) those without critical state [41]. The required parameters for the EVP models, satisfying the non-associated flow rule, ranges from 7 parameters [40] to 44 parameters [41]. A summary of the EVP models with the non-associated flow rule for different geomaterials is presented in Table 1.

**Table 1.** Summary of the EVP models using the non-associated flow rule for different geomaterials.

| Model | Material | Parameter | CSSM | Surface Shape |
|---|---|---|---|---|
| Model in this paper | Clay | 6 | Yes | NC |
| Zienkiewicz et al.[40] | Clay | 9 | Yes | NC |
| Kutter and Sathialingam [42] | Clay | 7 | Yes | C |
| Hickman and Gutierrez [43] | Chalk | 10 | Yes | NC |
| Cristescu [44] | Sand | 15 | No | C |
| Florea [45] | Concrete | 15 | No | C |
| Jin and Cristescu [46] | Rock | 32 | No | C |
| Maranini and Yamaguchi [41] | Granite | 44 | No | C |

Note: C, Circular; NC, Non-circular; CSSM, Critical state soil mechanics.

### 3.3. Model Parameters

There are six parameters in our model that need to be determined: the consolidation parameters [related to the gradient of the normal consolidation line $(\lambda)$ and the swelling line $(\kappa)$], the strength parameter [the slope of the critical state line $(M)$], the Poisson's ratio $(\nu)$, the state parameter, i.e., the void ratio $(e_N)$ at the unit mean pressure at any reference time $(\bar{t})$, and the creep parameter, i.e., the secondary compression index $(C_\alpha)$. Among these six parameters, five are similar to the MCC model parameters $(\lambda, \kappa, M, \nu, e_N)$. The details of how these parameters can be obtained were given by Roscoe and Burland [21]. The other parameter, $C_\alpha = \frac{\Delta e}{\Delta log(t)}$, can be obtained from either the oedometer test or the triaxial test. Here, $\Delta e$ is the change of the void ratio during the time change $\Delta log(t)$. In the literature, there are different definitions available for the secondary compression index (see Augustesen et al. [47]). The two most popular definitions are: (i) the void ratio based; and (ii) the strain based. In this paper, we use the first definition. Augustesen et al. [47] (see Figure 8, p. 141), presented a schematic diagram to calculate the secondary compression index. A similar definition can also be found in many soil mechanics textbooks.

### 3.4. Summary of the Basic Equations and the Assumptions Used in the Code

The equations used in the code and the finite element formulation do not have the same exact forms as those given in Sections 3.1–3.2. In this subsection, we present a summary of the derivation of the equations used in the code and we show how these equations can be obtained from the equations in the earlier sections, if proper assumptions are made. For example, the momentum (equilibrium) equation for the mixture, as used in finite element code, can be obtained if we write Equation (18) for the two components, add them up, assume that the inertial terms can be ignored, use the definition of the total stress tensor [Equation (19)], and the definition for the total density [Equation (11)], then we have

$$div\boldsymbol{\sigma} + \rho\boldsymbol{g} = \boldsymbol{0} \tag{48}$$

where $\boldsymbol{g}$ is the acceleration due to the gravity. If we substitute Equation (32) into Equation (48), and use the convention that a compressive pressure is positive, we have

$$div\left(\boldsymbol{\sigma}' + p_f\boldsymbol{I}\right) + \rho\boldsymbol{g} = \boldsymbol{0} \tag{49}$$

Similarly, if the conservation of mass equations, given by Equation (17), are re-written for the two components, for the cases of constant densities, i.e., when $\rho_R^s$ = constant and $\rho_R^f$ = constant, they become

$$-\frac{\partial\varphi}{\partial t} + div\left((1-\varphi)\,\boldsymbol{v}^s\right) = 0 \tag{50}$$

$$\frac{\partial\varphi}{\partial t} + div\left(\varphi\boldsymbol{v}^f\right) = 0 \tag{51}$$

Adding these two equations, we obtain

$$div\left[\varphi(\boldsymbol{v}^f - \boldsymbol{v}^s)\right] + div\boldsymbol{v}^s = 0 \tag{52}$$

In this equation, the term $\boldsymbol{v}_r = \varphi(\boldsymbol{v}^f - \boldsymbol{v}^s)$ is known as the "specific discharge" and is related to the relative velocity $\boldsymbol{w}$. We also notice that

$$div\,\boldsymbol{v}^s = -\frac{\partial\varepsilon_v}{\partial t} \tag{53}$$

[recall that $\varepsilon_v = tr\boldsymbol{\varepsilon}$ was defined as the dilatation (see Equation (6)), and $\dot{\varepsilon_v} = \frac{\partial(tr\boldsymbol{\varepsilon})}{\partial t} = \frac{\partial}{\partial t}\left(\frac{\partial u_i}{\partial x_i}\right) = div\boldsymbol{v}^s$]. Now, using Equations (53) and (30) in Equation (52), we obtain

$$div\left[\frac{\boldsymbol{K}}{\gamma_f}\left(\nabla p^f - \rho_f\boldsymbol{g}\right)\right] + \frac{\partial\varepsilon_v}{\partial t} = 0 \tag{54}$$

This expression is also known as the "storage equation" which is the basic equation for the consolidation theory ( see Bear and Bachmat [48], Equation 4.1.46).

## 4. Finite Element Solutions

The set of equations that need to be solved is: (i) the mass balance equation; (ii) the equilibrium equation; (iii) stress–strain relations; and (iv) strain–displacement relations. The unknown variables are the stresses, the strains, the displacements and the pore pressure. A summary of the equations and the unknown variables is presented in Table 2.

**Table 2.** Balance equations, constitutive relations and the unknown variables.

| Equations | | | Unknown Variables | |
|---|---|---|---|---|
| Name | Reference | Number | Variables | Number |
| Mass balance | Equation (54) | 1 | $p^f$ | 1 |
| Equilibrium | Equation (49) | 3 | $\boldsymbol{\sigma}'$ | 6 |
| Stress–strain | Equation (37) | 6 | $\boldsymbol{\varepsilon}$ | 6 |
| Strain–displacement | Equation (7) | 6 | $\boldsymbol{u}$ | 3 |
| Total equation | | 16 | Total unknowns | 16 |

Using the non-associated flow rule EVP model presented in this paper, we can see that there are enough equations to solve for the unknown variables. For any deformable porous media, similar expressions can be found in the work of Bear and Bachmat ([48], Chapter 4).

To solve these equations using finite element, we can use either "the strong form" or "the weak form". The details of these two techniques can be found in many finite element text books (Zienkiewicz et al. [49]). Even though both approaches provide similar results, considering the "continuity requirements" and the "symmetry of the stiffness matrix", the weak form solution is better compared to the strong form [49]. In this paper, we use the weak form approach. To obtain the weak form, the Galerkin weighted residual method [50] is introduced. Equations (48) and (54) are used to solve the problem. In the following section, a summary of the weak formulation is described.

*Weak Form Formulation*

Equilibrium equation

$$\boldsymbol{\mathcal{L}}^T\boldsymbol{\sigma} + \boldsymbol{b} = 0 \qquad \text{in } \Omega \tag{55}$$

Darcy's equation

$$\boldsymbol{w} = \frac{\boldsymbol{K}}{\gamma_w}\left(\nabla p_f - \rho_f \boldsymbol{g}\right) \qquad \text{in } \Omega \tag{56}$$

Continuity equation

$$\nabla^T\boldsymbol{w} + \boldsymbol{m}^T\dot{\varepsilon_v} = 0 \qquad \text{in } \Omega \tag{57}$$

Effective stress relation

$$\boldsymbol{\sigma} = \boldsymbol{\sigma}' + \boldsymbol{m}p_f \tag{58}$$

In Equations (55)–(58), $\boldsymbol{\sigma}$ and $\boldsymbol{\sigma}'$are the total stress and the effective stress, respectively, $\boldsymbol{b}$ is the body force, $\boldsymbol{\mathcal{L}}^T$is the differential operator, $\nabla^T$is the divergence operator, $\nabla$ is the gradient operator, $p_f$ is the pore pressure, $\boldsymbol{w}$ is the superficial velocity, $\boldsymbol{K}$ is the coefficient of permeability, $\gamma_w$ is the unit weight of the fluid, $\boldsymbol{b}_f$ is the body force vector for the fluid, and $\dot{\boldsymbol{\varepsilon}}$ is the volumetric strain rate of soil. The general expression of $\boldsymbol{\mathcal{L}}^T$, $\nabla$, $\boldsymbol{b}$, $\boldsymbol{b}_f$ $\boldsymbol{m}$, $\boldsymbol{w}$ and $\boldsymbol{K}$ in the above Equations are

$$\boldsymbol{\mathcal{L}}^T = \begin{bmatrix} \frac{\partial}{\partial x} & 0 & 0 & \frac{\partial}{\partial y} & 0 & \frac{\partial}{\partial z} \\ 0 & \frac{\partial}{\partial y} & 0 & \frac{\partial}{\partial x} & \frac{\partial}{\partial z} & 0 \\ 0 & 0 & \frac{\partial}{\partial z} & 0 & \frac{\partial}{\partial y} & \frac{\partial}{\partial x} \end{bmatrix} \tag{59}$$

$$\nabla = \left[\frac{\partial}{\partial x}, \frac{\partial}{\partial y}, \frac{\partial}{\partial z}\right]^T \tag{60}$$

$$\boldsymbol{b} = \left[b_x, b_y, b_z\right]^T \tag{61}$$

$$\boldsymbol{b}_f = [0, \gamma_w, 0]^T \tag{62}$$

$$\boldsymbol{m} = [1,1,1,0,0,0]^T \tag{63}$$

$$\boldsymbol{w} = [w_x, w, w_z]^T \tag{64}$$

$$\boldsymbol{K} = \begin{bmatrix} K_x & 0 & 0 \\ 0 & K_y & 0 \\ 0 & 0 & K_z \end{bmatrix} \tag{65}$$

Assuming that the soil is subjected to the traction $\boldsymbol{t}$, the following conditions can be written. Initial conditions:

$$\boldsymbol{u}(t=0) = \boldsymbol{u}_0 \qquad \text{in } \Omega \tag{66}$$

$$p_f(t=0) = p_{f_0} \qquad \text{in } \Omega \tag{67}$$

Boundary conditions:
for the soil,

$$\boldsymbol{\sigma}.\boldsymbol{n} = \boldsymbol{t} \qquad \text{on } \Gamma_t \tag{68}$$

$$\boldsymbol{u} = \boldsymbol{u}_0 \qquad \text{on } \Gamma_u \tag{69}$$

for the fluid,

$$p_f = p_{f_0} \qquad \text{on } \Gamma_{p_f} \tag{70}$$

$$\boldsymbol{v}.\boldsymbol{n} = 0 \qquad \text{on } \Gamma_v \tag{71}$$

In Equations (66)–(71), $\boldsymbol{u}_0$ is the initial displacement, $p_{f_0}$ is the initial fluid pressure, $\boldsymbol{n}$ is the unit normal vector, $\Gamma_t$, $\Gamma_u$, $\Gamma_{p_f}$and $\Gamma_v$ are the respective segment of the boundary, and the subscripts have similar meanings as described above. The surface traction vector is defined as

$$\boldsymbol{t} = \left[t_x, t_y, t_z\right]^T \tag{72}$$

$$t_x = \sigma_x n_x + \tau_{xy} n_y + \tau_{xz} n_z \tag{73}$$

$$t_y = \tau_{xy} n_x + \sigma_y n_y + \tau_{yz} n_z \tag{74}$$

$$t_z = \tau_{xz} n_x + \tau_{yz} n_y + \sigma_z n_z \tag{75}$$

In Equations (66)–(71), two types of boundary conditions are used: (i) the essential boundary condition (EBC) or the Dirichlet boundary condition; and (ii) the natural boundary condition (NBC) or the Neumann boundary condition. For the EBC, the primary variables in the domain are known, while, for the NBC, the differential form of the primary variables are prescribed.

Assuming that soil is isotropic, the stress tensor $\boldsymbol{\sigma}$, the stain tensor $\boldsymbol{\varepsilon}$ and the displacement vector $\boldsymbol{u}$ can be expresses as

$$\boldsymbol{\sigma} = \left[\sigma_{xx}, \sigma_{yy}, \sigma_{zz}, \sigma_{xy}, \sigma_{yz}, \sigma_{zx}\right]^T \tag{76}$$

$$\boldsymbol{\varepsilon} = \left[\varepsilon_{xx}, \varepsilon_{yy}, \varepsilon_{zz}, \varepsilon_{xy}, \varepsilon_{yz}, \varepsilon_{zx}\right]^T \tag{77}$$

$$\boldsymbol{u} = \left[u_{xx}, u_{yy}, u_{zz}\right]^T \tag{78}$$

To obtain the weak formulation for the equilibrium equation, basic necessary steps are: (i) replacing Equation (58) in Equation (55) and multiplying the resulting equation with an arbitrary function, which removes the essential boundary condition or the Dirichlet boundary condition [51]; (ii) integrating the system over the domain; and (iii) applying the Green–Gauss theorem [50] and integrating by parts .

$$\int_\Omega [\boldsymbol{B}_u]^T \dot{\boldsymbol{\sigma}}' \mathrm{d}\Omega + \int_\Omega [\boldsymbol{B}_u]^T \boldsymbol{m} \dot{p}_f \mathrm{d}\Omega = \int_\Omega [\boldsymbol{N}_u]^T \dot{\boldsymbol{b}} \mathrm{d}\Omega + \oint_\Gamma [\boldsymbol{N}_u]^T \dot{\boldsymbol{T}} d\Gamma \tag{79}$$

Now, $\dot{\boldsymbol{\sigma}}'$ (see Equation (37)) can be written as

$$\dot{\boldsymbol{\sigma}'} = \boldsymbol{D}(\boldsymbol{B}_u \dot{\boldsymbol{u}}^n) - \boldsymbol{D}\dot{\boldsymbol{\varepsilon}}^{vp} \tag{80}$$

where, in Equation (37), $\dot{\boldsymbol{\epsilon}} = \boldsymbol{B}_u \dot{\boldsymbol{u}}^n$ and in Equation (79), $\dot{p}_f = \boldsymbol{N}_p \dot{\boldsymbol{p}}_f^n$ [Zienkiewicz et al. [49]]. The compact form of Equation (79) can be written as

$$\boldsymbol{K}_E \dot{\boldsymbol{u}}^n + \boldsymbol{I}_E \dot{\boldsymbol{p}}_f^n = \boldsymbol{F}_E \quad (81)$$

Where

$$\boldsymbol{K}_E = \int_\Omega [\boldsymbol{B}_u]^T \boldsymbol{D} \boldsymbol{B}_u \mathrm{d}\Omega \quad (82)$$

$$\boldsymbol{I}_E = \int_\Omega [\boldsymbol{B}_u]^T \boldsymbol{m} \boldsymbol{N}_p \mathrm{d}\Omega \quad (83)$$

$$\boldsymbol{F}_E = \int_\Omega [\boldsymbol{N}_u]^T \dot{\boldsymbol{b}} \mathrm{d}\Omega + \oint_\Gamma [\boldsymbol{N}_u]^T \dot{\boldsymbol{T}} d\Gamma + \int_\Omega [\boldsymbol{B}_u]^T \boldsymbol{D} \dot{\boldsymbol{\varepsilon}}^{vp} \mathrm{d}\Omega \quad (84)$$

$$\boldsymbol{D} = [(\boldsymbol{D}^e \boldsymbol{C})^{-1} + \theta_d \Delta t_n \boldsymbol{H}_n]^{-1} \quad (85)$$

$$\boldsymbol{H}_n = \left(\frac{\partial \dot{\boldsymbol{\varepsilon}}^{vp}}{\partial \boldsymbol{\sigma}'}\right)_n \quad (86)$$

$$\boldsymbol{B}_u = \boldsymbol{\mathcal{L}} \boldsymbol{N}_u \quad (87)$$

$$\boldsymbol{N}_u = \begin{bmatrix} N_{u1} & 0 & 0 & \dots & N_{un} & 0 & 0 \\ 0 & N_{u1} & 0 & \dots & 0 & N_{un} & 0 \\ 0 & 0 & N_{u1} & \dots & 0 & 0 & N_{un} \end{bmatrix} \quad (88)$$

$$\boldsymbol{N}_p = [N_{p1} \quad \dots \quad N_{pn}] \quad (89)$$

$$\boldsymbol{m} = [1,1,1,0,0,0]^T \quad (90)$$

$$\boldsymbol{b} = [b_x, b_y, b_z]^T \quad (91)$$

$$\boldsymbol{\mathcal{L}} = \begin{bmatrix} \frac{\partial}{\partial x} & 0 & 0 \\ 0 & \frac{\partial}{\partial y} & 0 \\ 0 & 0 & \frac{\partial}{\partial z} \\ \frac{\partial}{\partial y} & \frac{\partial}{\partial x} & 0 \\ 0 & \frac{\partial}{\partial z} & \frac{\partial}{\partial y} \\ \frac{\partial}{\partial z} & 0 & \frac{\partial}{\partial x} \end{bmatrix} \quad (92)$$

Similar formulation for Equation (57) can be written as

$$[\boldsymbol{I}_E]^T \dot{\boldsymbol{u}}^n + \boldsymbol{H}_E \boldsymbol{p}_f^n = \boldsymbol{Q}_E \quad (93)$$

where

$$[\boldsymbol{I}_E]^T = \int_\Omega [\boldsymbol{N}_p]^T \boldsymbol{m}^T \boldsymbol{B}_u \mathrm{d}\Omega \quad (94)$$

$$\boldsymbol{H}_E = -\int_\Omega [\boldsymbol{B}_p]^T \frac{\boldsymbol{K}}{\gamma_w} \boldsymbol{B}_p \mathrm{d}\Omega \quad (95)$$

$$\boldsymbol{Q}_E = -\int_\Omega [\boldsymbol{B}_p]^T \frac{\boldsymbol{K}}{\gamma_w} \boldsymbol{b}_f \mathrm{d}\Omega - \oint_\Gamma [\boldsymbol{N}_p]^T \boldsymbol{w} d\Gamma \quad (96)$$

$$\boldsymbol{B}_p = \nabla \boldsymbol{N}_p \quad (97)$$

$$\nabla = \left[\frac{\partial}{\partial x}, \frac{\partial}{\partial y}, \frac{\partial}{\partial z}\right]^T \tag{98}$$

Now, combining Equations (81) and (93), we obtain

$$\begin{bmatrix} \boldsymbol{K}_E & \boldsymbol{I}_E \\ [\boldsymbol{I}_E]^T & 0 \end{bmatrix} \begin{bmatrix} \dot{\boldsymbol{u}}^n \\ \dot{\boldsymbol{p}}_f^n \end{bmatrix} + \begin{bmatrix} 0 & 0 \\ 0 & \boldsymbol{H}_E \end{bmatrix} \begin{bmatrix} \boldsymbol{u}^n \\ \boldsymbol{p}_f^n \end{bmatrix} = \begin{bmatrix} \boldsymbol{F}_E \\ \boldsymbol{Q}_E \end{bmatrix} \tag{99}$$

Equation (99) represent the element matrix. To obtain the global matrix, Equation (99) needs to be summed over a number of elements as follows

$$\boldsymbol{K}_G = \sum_{n=1}^{n} \boldsymbol{K}_E \tag{100}$$

$$\boldsymbol{I}_G = \sum_{n=1}^{n} \boldsymbol{I}_E \tag{101}$$

$$[\boldsymbol{I}_G]^T = \sum_{n=1}^{n} [\boldsymbol{I}_E]^T \tag{102}$$

$$\boldsymbol{H}_G = \sum_{n=1}^{n} \boldsymbol{H}_E \tag{103}$$

$$\boldsymbol{F}_G = \sum_{n=1}^{n} \boldsymbol{F}_E \tag{104}$$

$$\boldsymbol{Q}_G = \sum_{n=1}^{n} \boldsymbol{Q}_E \tag{105}$$

In the time interval $\Delta t_n = t_{n+1} - t_n$, the viscoplastic strain rate increment $\Delta \boldsymbol{\varepsilon}_n^{vp}$ can be defined as [7,8,52]

$$\Delta \boldsymbol{\varepsilon}_n^{vp} = \Delta t_n \left[(1-\theta_d)\dot{\boldsymbol{\varepsilon}}_n^{vp} + \dot{\boldsymbol{\varepsilon}}_{n+1}^{vp}\right] \tag{106}$$

where $\theta_d = 0$, denotes the "fully explicit" Euler time integration scheme (or the forward difference method) and $\theta_d = 1$ indicates the "fully implicit" Euler time integration scheme (or the backward difference method). For $\theta_d \geq 0.5$, the integration scheme is unconditionally stable and, when $\theta_d < 0.5$, the integration process is conditionally stable. The integration scheme for $\theta_d = 0.5$ is the "implicit trapezoidal" scheme or the Crank–Nicolson rule.

The viscoplastic strain rate increment $\dot{\boldsymbol{\varepsilon}}_{n+1}^{vp}$ at time step ($n$ +1) in Equation (106) can also be written as follows [8]

$$\dot{\boldsymbol{\varepsilon}}_{n+1}^{vp} = \dot{\boldsymbol{\varepsilon}}_n^{vp} + \boldsymbol{H}_n \Delta \boldsymbol{\sigma}_n \tag{107}$$

where $\Delta \boldsymbol{\sigma}_n$ is the stress change at any time interval $\Delta t_n = t_{n+1} - t_n$ and $\boldsymbol{H}_n$ (given in Appendix D) is the derivative of the viscoplastic strain-rate with respect to time at the $n$th time step.

Combining Equations (106) and (107), the above can be written as follows [8]

$$\Delta \boldsymbol{\varepsilon}_n^{vp} = \Delta t_n \dot{\boldsymbol{\varepsilon}}_n^{vp} + \boldsymbol{C}_n \Delta \boldsymbol{\sigma}_n \tag{108}$$

Where,

$$\boldsymbol{C}_n = \theta_d \boldsymbol{H}_n \Delta t_n \tag{109}$$

$$\Delta \boldsymbol{\sigma}_n = \boldsymbol{D}\left(\boldsymbol{B}^n \Delta \boldsymbol{u}^n - \left(\Delta t_n \dot{\boldsymbol{\varepsilon}}_n^{vp} + \boldsymbol{C}_n \Delta \boldsymbol{\sigma}_n\right)\right) \tag{110}$$

## 5. Model Verification and Discussion

We investigated the performance of our EVP model in a variety of experimental applications. These applications included the consolidated undrained triaxial compression test and the extension test, the consolidated drained compression test, the over-consolidation ratio test, the confining pressure effect, the strain rate test, the creep test, and the relaxation test. In this respect, we considered both natural clay and laboratory obtained reconstituted clay. We specifically investigated the Osaka clay [53], the San Francisco Bay Mud (SFBM) clay [54], the Kaolin clay [55], and the Hong Kong Marine Deposit (HKMD) clay [56]. Clay properties are presented in Table 3 and the triaxial loading configuration is shown in Figure 4.

**Table 3.** Model parameters.

| Parameters | Osaka Clay [53] | SFBM Clay [54] | Kaolin Clay [55] | HKMD Clay [56] |
|---|---|---|---|---|
| $\lambda$ | 0.36 | 0.37 | 0.15 | 0.20 |
| $\kappa$ | 0.047 | 0.054 | 0.018 | 0.045 |
| $M_{b=0}$ | 1.28 | 1.40 | 1.25 | 1.26 |
| $M_{b=1}$ | — | — | 0.95 | 0.89 |
| $\nu$ | 0.3 | G = 23520 kPa | 0.30 | 0.30 |
| $e_N$ | 3.54 | 3.17 | 1.51 | 2.18 |
| $C_\alpha$ | 0.0327 | 0.053 | 0.014 | 0.106 |

$M_{b=0}$ and $M_{b=1}$ are for the triaxial compression and the extension test respectively.

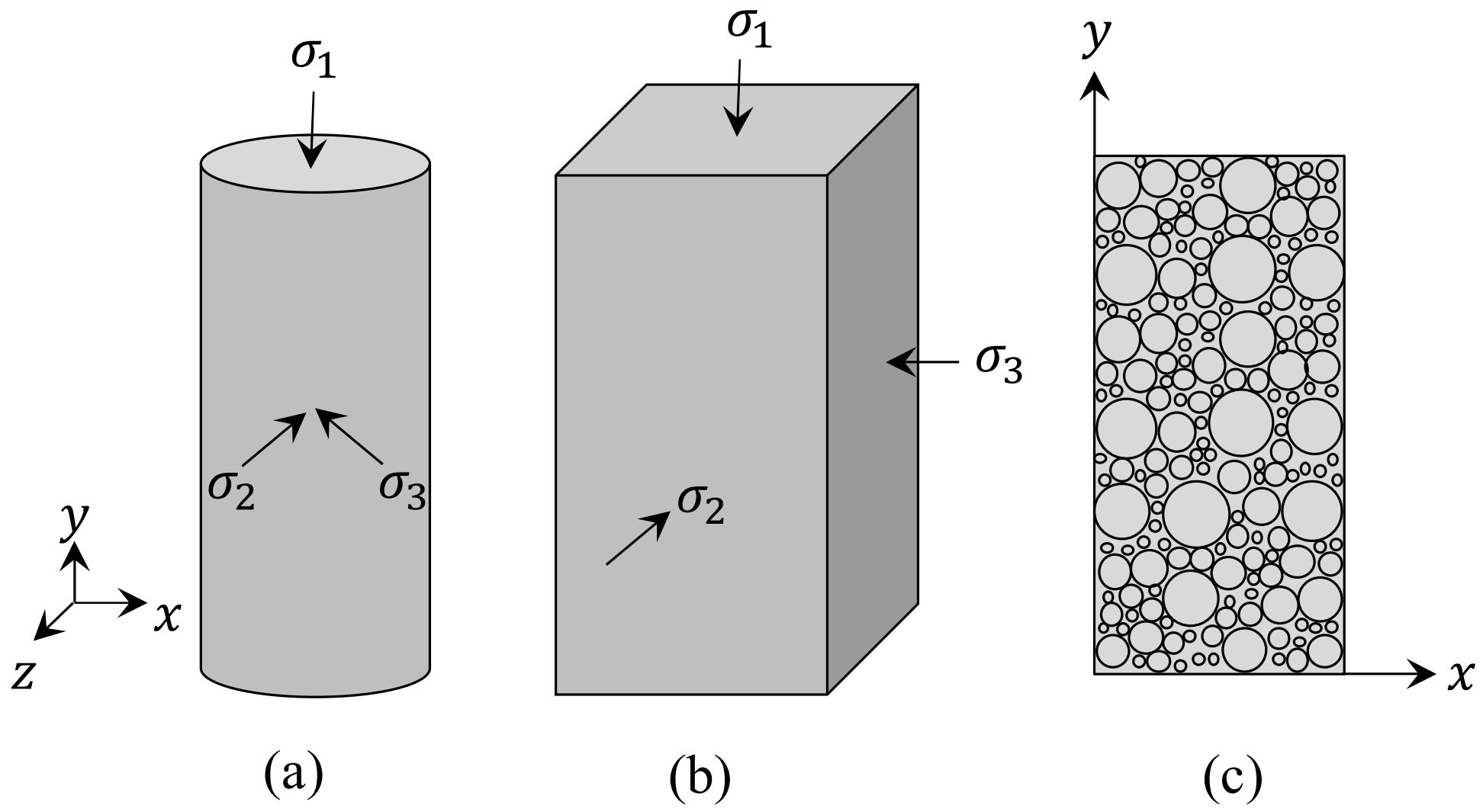

**Figure 4.** The stress condition in triaxial compression/extension tests (**a**) cylindrical sample, (**b**) block sample, (**c**) axisymmetric/biaxial representation of solid and fluids in the sample.

## *5.1. Simulation of the Undrained Triaxial Test on the Osaka Clay*

Adachi et al. [53] presented undrained triaxial compression test for the natural Osaka clay. This is a moderately sensitive clay (sensitivity, $S_t$ = 14.50). The water content, the liquid limit and the plastic limit of this clay are 65.00–72.00%, 69.20–75.10% and 24.50–27.30%, respectively. Its specific gravity is 2.62–2.70. The particle size distribution of this clay consists of the clay fraction of 44.00%, the silt fraction of 49.00% and the sand fraction of 7.00%. The strain rate during the test was $1 \times 10^{-2}$ %/minute. From the experimental results, it is evident that the Osaka clay shows strain softening behavior, which was captured using the new EVP model presented in this paper.

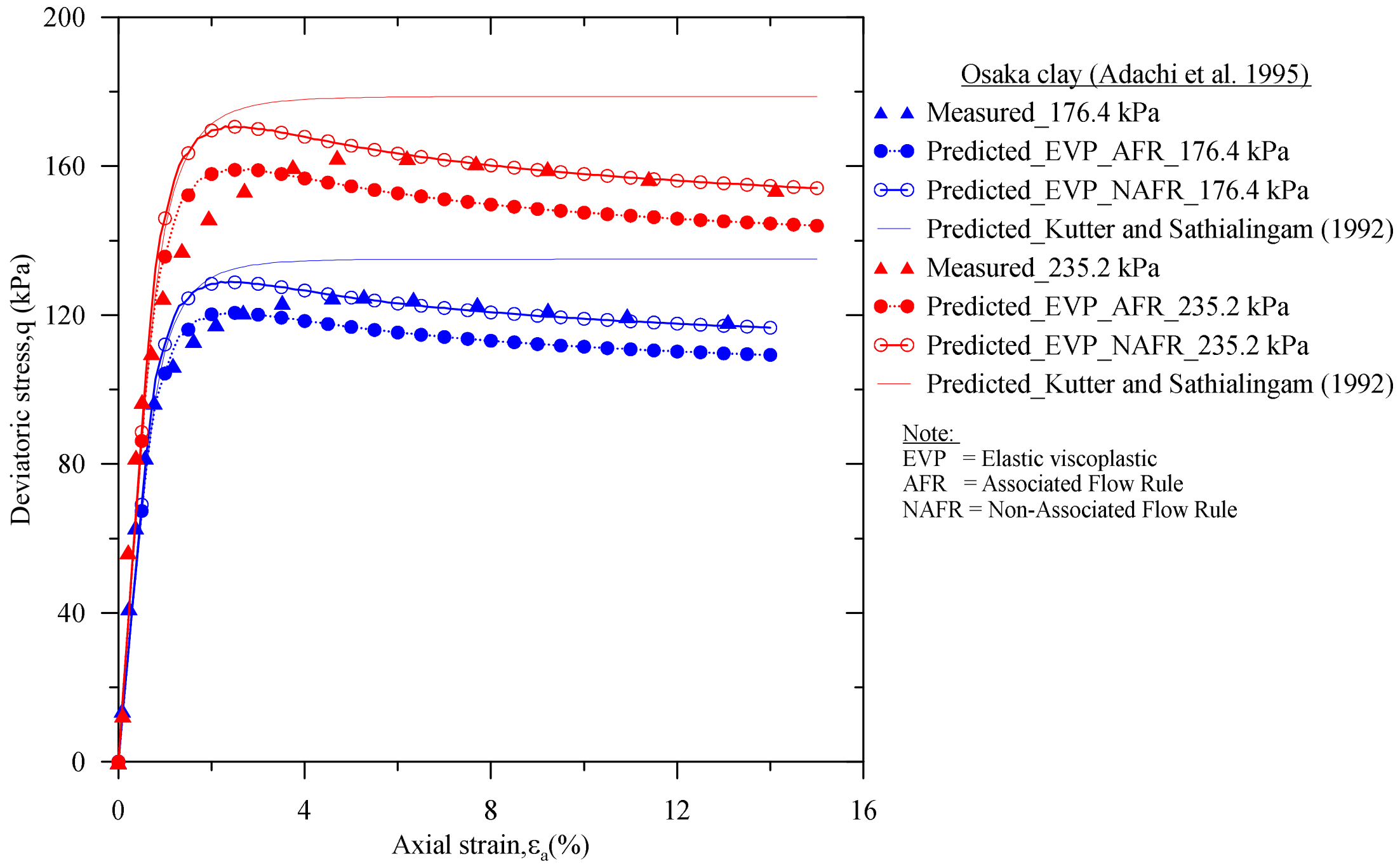

(**a**)

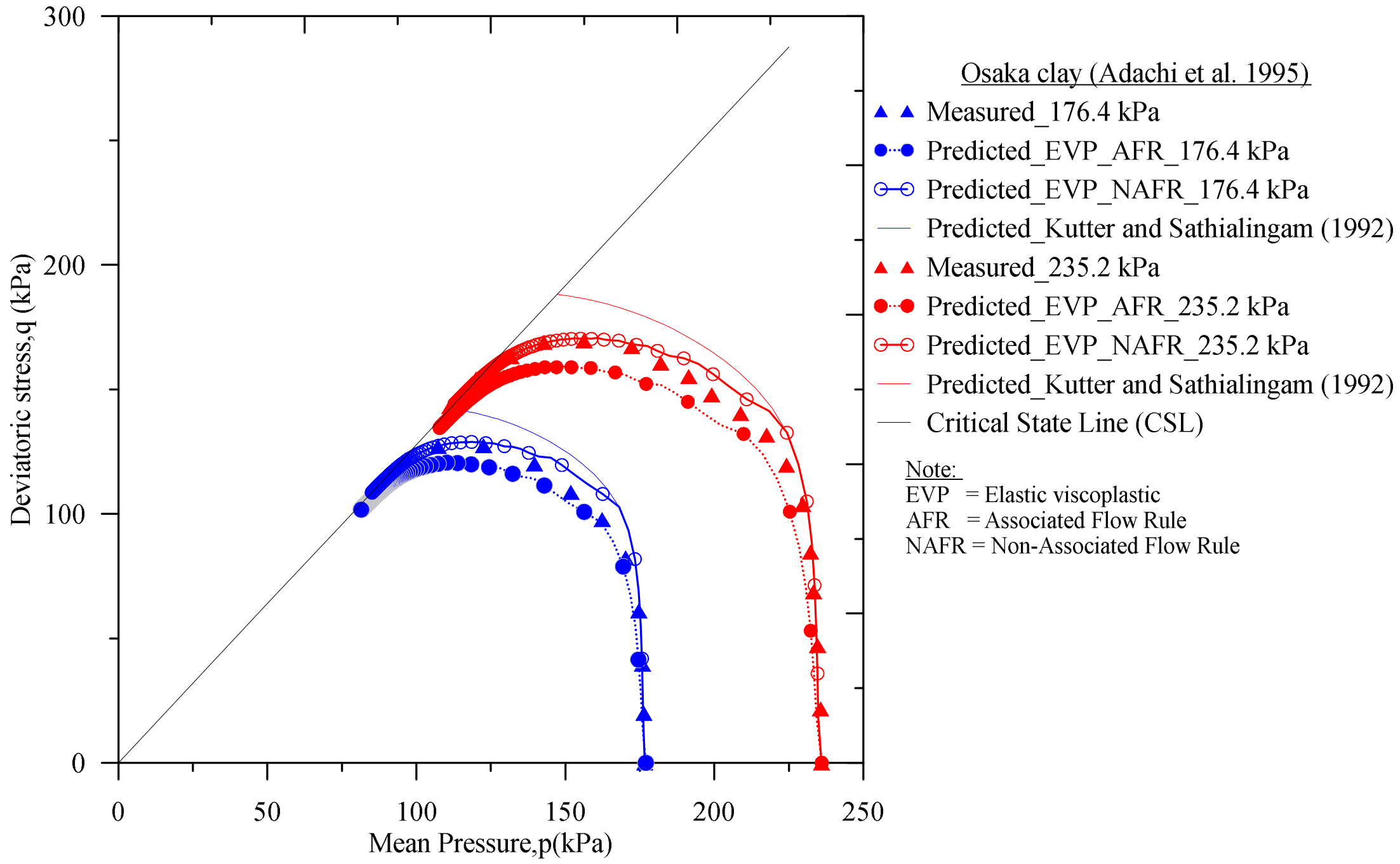

(**b**)

**Figure 5.** Comparison of the measured and the predicted consolidated undrained triaxial compression tests results for the Osaka clay: (**a**) the deviatoric stress vs. axial strain; and (**b**) the stress path.

In Figure 5, the experimental results for the Osaka clay for two different mean pressures (176.4 kPa and 235.2 kPa) are compared with the non-associated flow rule type EVP model, and the EVP model developed by Kutter and Sathialingam [42]. In Figure 5-a, we can see that, for the mean pressure of 235.2 kPa, the new EVP model and the Kutter and Sathialingam [42] EVP model predictions were identical up to an axial strain of 1.87%. After this point, their model over-predicted the observed experimental results. After an axial strain of 14%, the over-prediction in their model was 16%, whereas the new EVP model captured the strain softening. On the other hand, at an axial strain of 14%, the under-prediction in the new EVP model was 5.20%. For small strain cases (axial strains of 1–3.75%), an over-prediction was observed in both EVP models. Jiang et al. [57] reported that, at small strains, the over-prediction of the nonlinear responses of clay are expected. Using the hysteretic response equation [58], such issues can be resolved; however, this requires additional model parameters. To make the present EVP model formulation simple, and to limit the number of model parameters, the hysteretic response equation concept was not introduced here. A similar trend was also observed for the mean pressure of 176.4 kPa.

In Figure 5-b, the stress paths obtained from the associated flow rule, the non-associated flow rule EVP models, and the Kutter and Sathialingam model are compared with the data for the Osaka clay. It is evident that, after the deviatoric stress reached the peak, the attributed stress paths gradually followed the "narrow region". This phenomenon indicates that the critical state concept is applicable to the natural soft clay, even at large strains [53]. It was observed that the new non-associated flow rule EVP model could capture the "narrow region", whereas there was marginal under-prediction in the associated flow rule EVP model.

### *5.2. Simulation of the Undrained Triaxial Stress Relaxation Test on the San Francisco Bay Mud Clay*

Lacerda [54] presented the results of undrained triaxial stress relaxation tests for the undisturbed San Francisco Bay Mud clay. In this paper, only SR-I-5 test data are presented and compared with the EVP model predictions. The sample was isotropically consolidated to a pressure of 78.4 kPa. Basic properties of this clay include the specific gravity = 2.66–2.75; the liquid limit = 88.4–90%; the plastic limit = 35–44 %; the plasticity index = 45–55%; and the moisture content = 88–93%. Kaliakin and Dafalias [59] reported that the initial void ratio $(e_0)$ and its corresponding mean pressure $(p_0)$ for the San Francisco Bay Mud are 1.30 and 156.9 kPa, respectively. The details of the undrained triaxial stress relaxation tests on the SFBM clay are presented in Table 4.

**Table 4.** Stress relaxation test on the SFBM clay.

| **Phase** | **1** | **2** | **3** | **4** | **5** | **6** | **7** | **8** |
|---|---|---|---|---|---|---|---|---|
| Test | Shear | Rel. | Shear | Rel. | Shear | Rel. | Shear | Rel. |
| $\dot{\varepsilon}_a$ | 1.5 | 0 | 1.5 | 0 | 0.0162 | 0 | 0.00081 | 0 |
| Time | 0.25 | 3070 | 1.28 | 1320 | 101.24 | 2700 | 1679 | 8370 |
| $\varepsilon_a$(%) | 0-0.38 | 0.38 | 0.38-2.3 | 2.3 | 2.3-3.94 | 3.94 | 3.94-5.3 | 5.3 |

Rel. = Relaxation, $\dot{\varepsilon}_a$ = (%/ min), Time = minutes.

In Figure 6, the measured data for the San Francisco Bay Mud clay are compared with the numerical simulations for the associated flow rule and the non-associated flow rule EVP models. In addition, our results are also compared with the EVP models of Kaliakin and Dafalias [59] and Kutter and Sathialingam [42]. It is evident that the non-associated flow rule prediction was close to the laboratory results.

### *5.3. Simulation of the Consolidated Undrained Triaxial Tests on the Kaolin Clay*

Herrmann et al. [55] presented the effect of over-consolidation ratio (OCR) on the Kaolin clay for the undrained triaxial compression test and the extension test. This clay is a mixture of the Kaolin (Snow-Cal 50) and a 5.0% Bentonite by weight, which was prepared in the laboratory. The Kaolin clay properties are: the specific gravity = 2.64, the liquid limit = 47.0%, and the plastic limit = 20.0%. The sample was isotropically consolidated to a pressure of 392.2 kPa; to achieve OCR = 1, 2, 4 and 6, the confining pressure was changed accordingly. For the OCR = 1, the corresponding initial void ratio $(e_0)$ is 0.613. The deviatoric stress versus the axial strain predictions for the consolidated undrained triaxial compression test (OCR = 1, 2, 4 and 6) and the extension test (OCR = 1 and 2) were compared with the experimental results, as shown in Figure 7. For the normally consolidated clay (OCR=1), it was observed that the new EVP model captured the stress–strain responses well before the deviatoric stress reached its peak, but then slightly under-predicted them (at 14% strain, the under-prediction was only 1.05 %). In Figure 7a, it is evident that, when OCR = 2, 4 and 6, before reaching the peak deviatoric stress, the model over-predicted but then exhibited small amount of under-prediction. In the consolidated undrained extension tests, a similar trend was observed. When OCR = 1 and 2, we compared the experimental results with the predicted responses of the pore-water pressure for the consolidated undrained triaxial compression test and the consolidated undrained triaxial extension tests, as shown in Figure 7b. For the normally consolidated clay, predictions for both the compression test and the extension test were satisfactory. For the OCR = 2, in the triaxial compression test, the new model slightly over-predicted before the maximum pore pressure was reached and, in the triaxial extension, the negative pore pressure was well captured with a small degree of under-prediction. A similar pattern was observed for the stress path, as shown in Figure 7c.

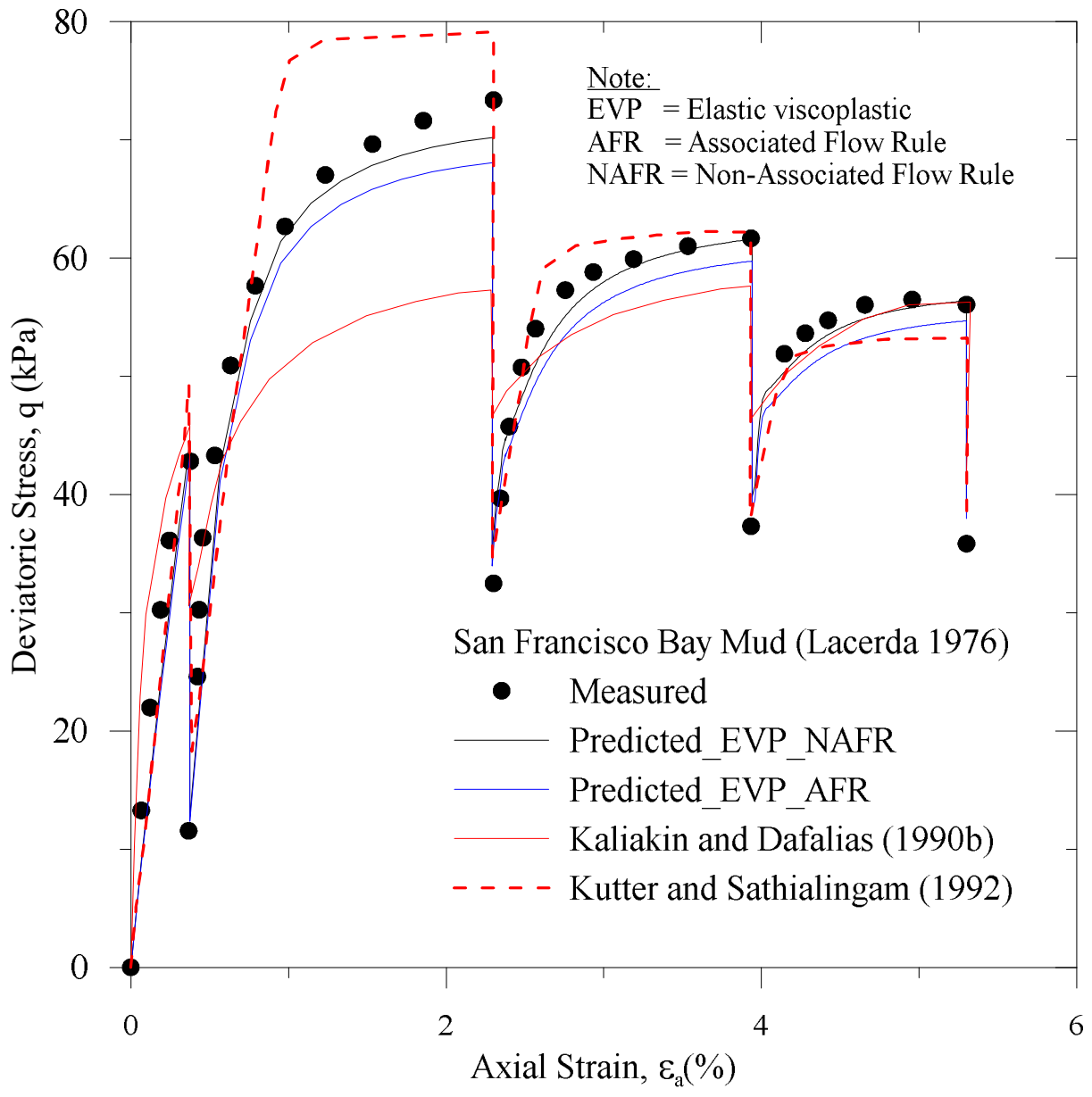

(**a**)

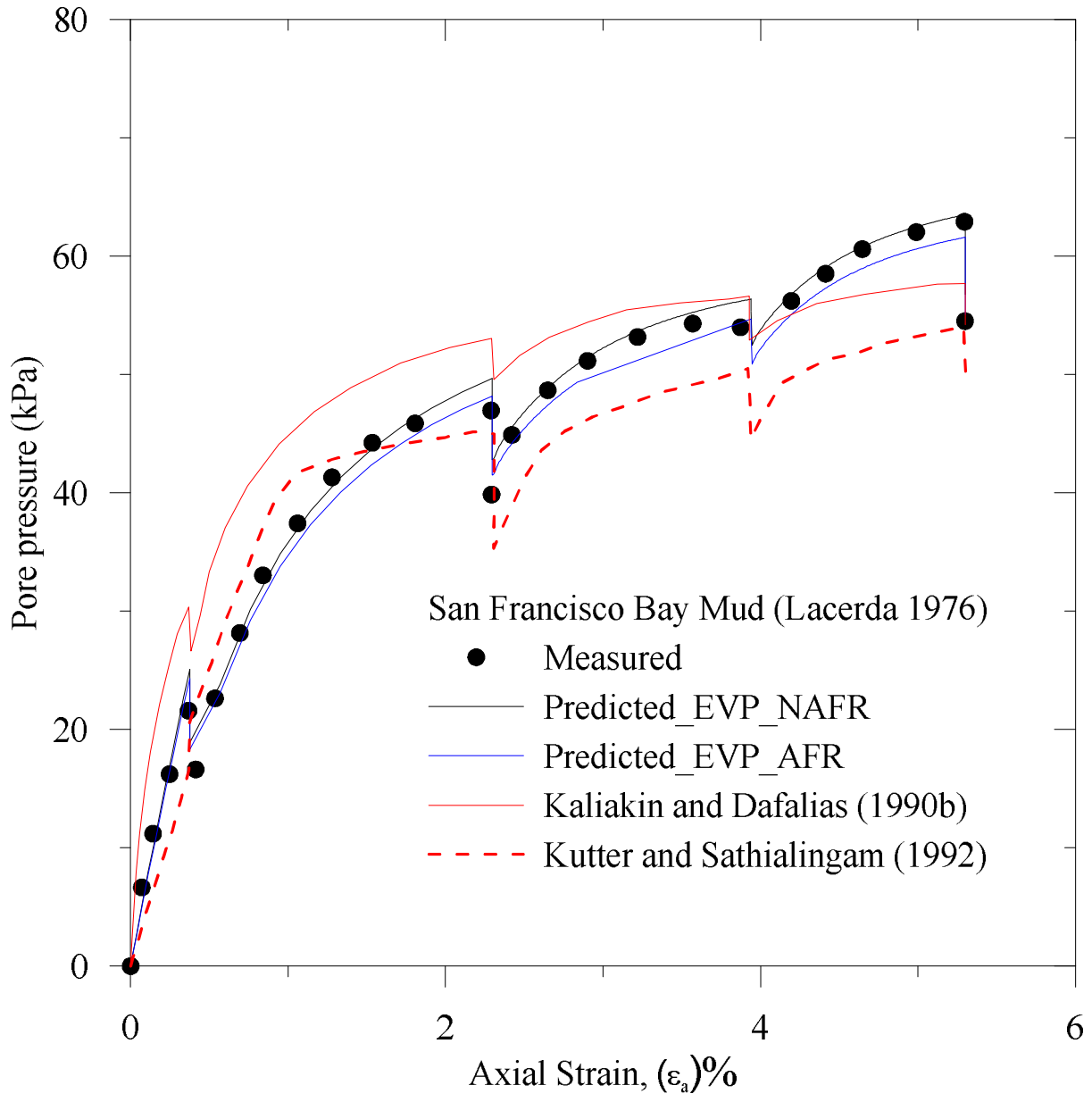

(**b**)

**Figure 6.** Comparison of the measured and the predicted undrained triaxial tests for the stage-changed axial strain rate combined with stress relaxation on the SFBM clay: (**a**) the deviatoric stress vs. axial strain; and (**b**) the pore pressure vs. axial strain.

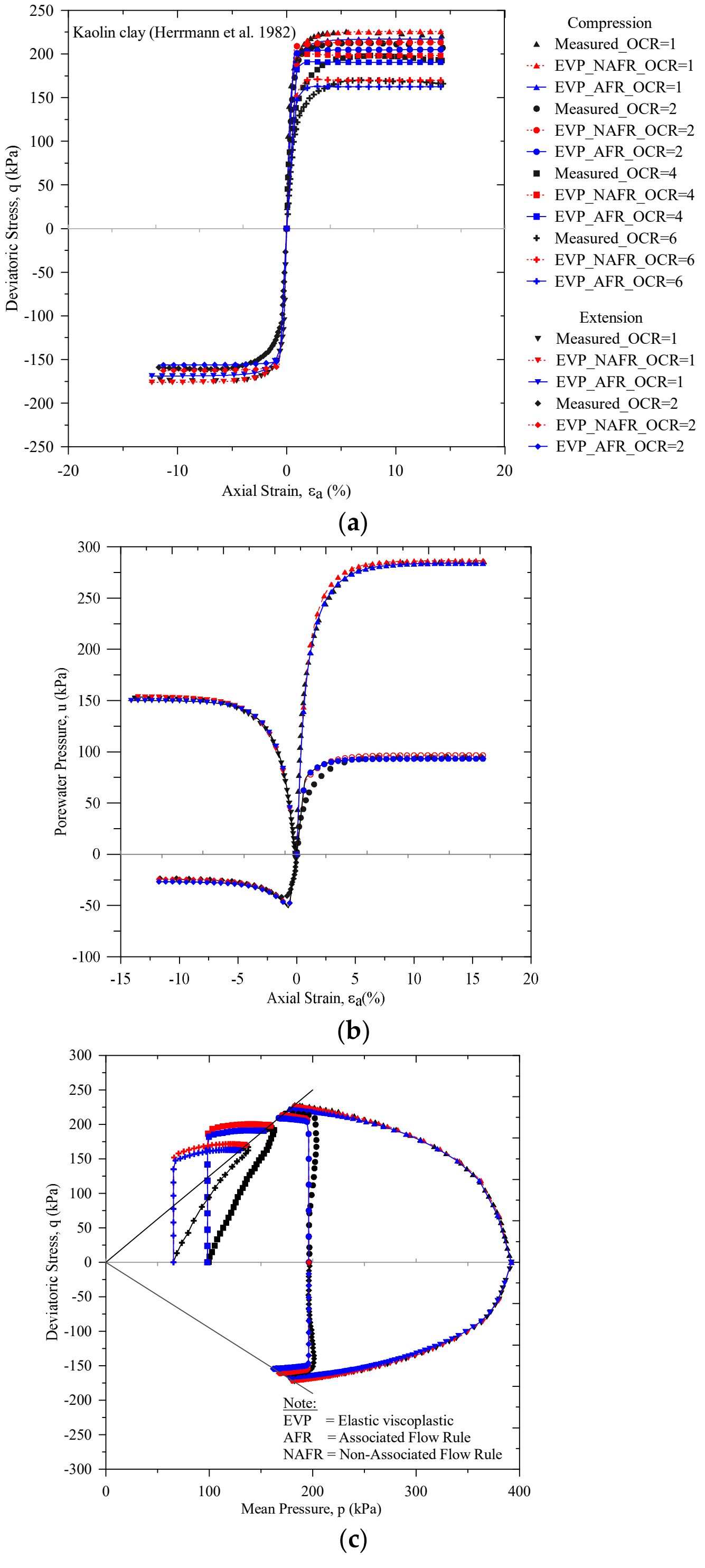

**Figure 7.** Comparison of the measured and the predicted consolidated undrained triaxial compression and the extension tests on the Kaolin clay: (**a**) the deviatoric stress vs. the axial strain; (**b**) the pore-water pressure vs. the axial strain; and (**c**) the stress path.

### *5.4. Simulation of the Consolidated Drained Tests on the Hong Kong Marine Deposit Clay*

The Hong Kong Marine Deposit clay [see Yin and his co-workers [56]] was used to study the isotropically consolidated drained test. This is a reconstituted soft to very soft illitic silty clay, which has been extensively studied.

The properties of the Hong Kong Marine Deposit (HKMD) clay are: the specific gravity = 2.66, the liquid limit = 60%, the plastic limit = 28%, the plasticity index = 32%, and the moisture content = 51.7%. The predictions of the new model for the normally consolidated drained triaxial test on the HKMD clay are shown in Figure 8a for the mean pressures of 300 kPa and 400 kPa. It is evident that the present model captured the deviatoric stress versus axial strain response very well (Figure 8). The volumetric response and the stress paths are presented in Figure 8b, c, which were also satisfactory.

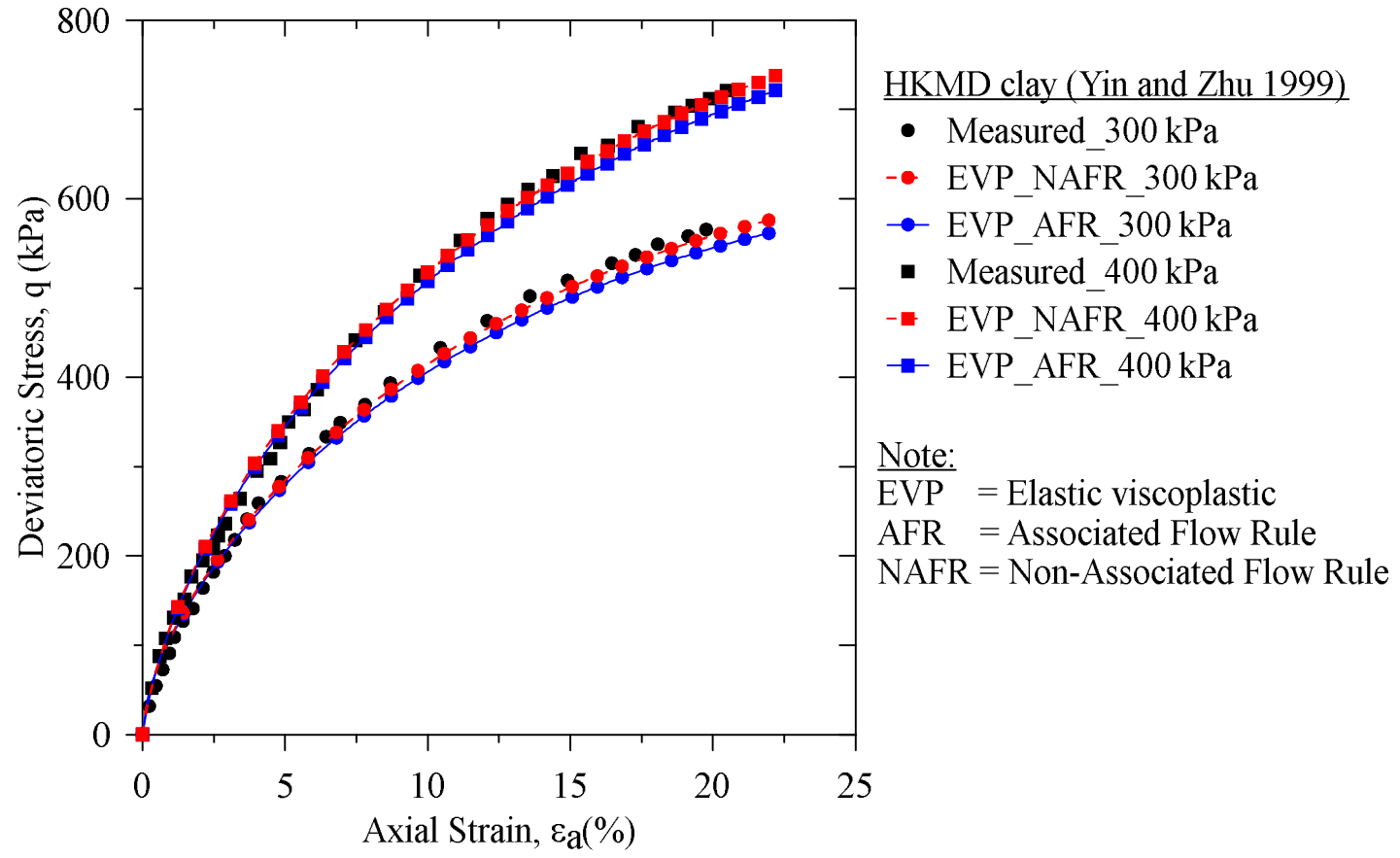

(**a**)

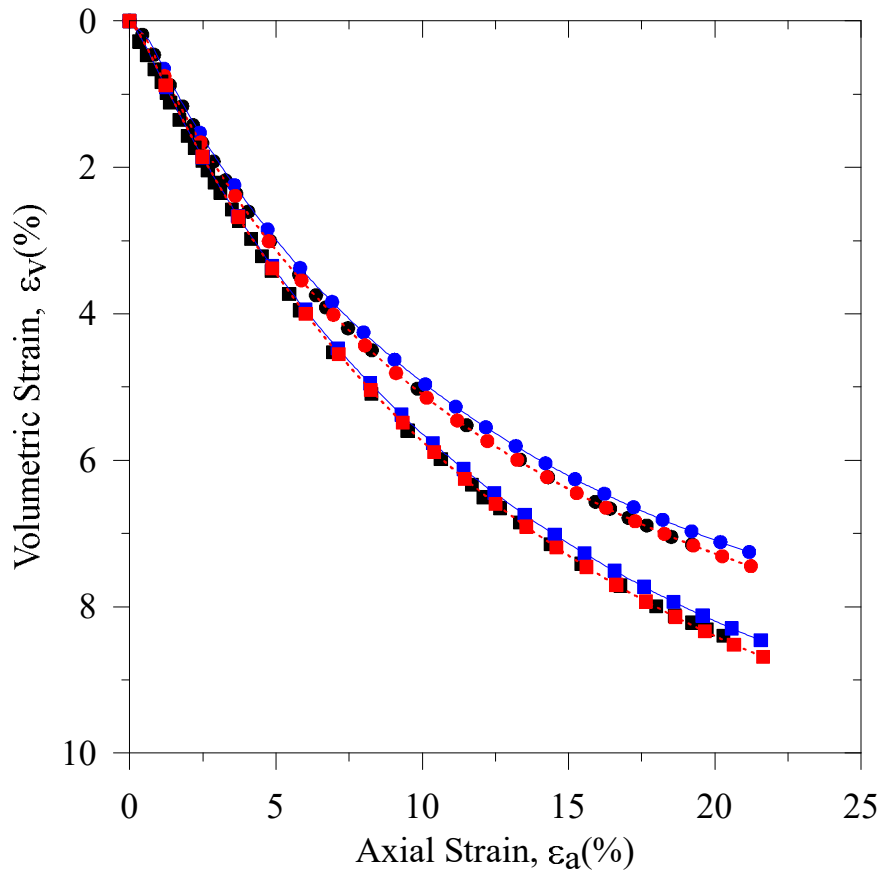

(**b**)

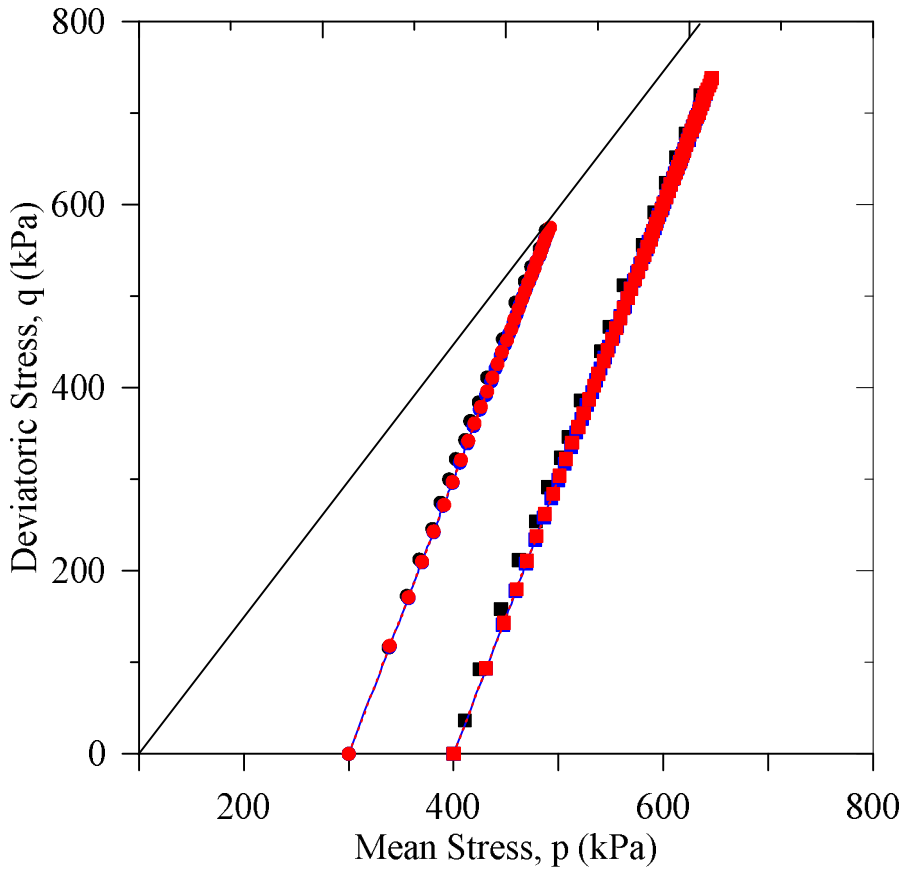

(**c**)

**Figure 8.** Comparison of the measured and the predicted consolidated drained triaxial compression tests on the HKMD clay: (**a**) the deviatoric stress vs. the axial strain; (**b**) the volumetric strain vs. the axial strain; and (**c**) the stress path.

## 6. Perturbation Analysis

We performed a perturbed analysis for each model parameter for the Osaka clay and the Kaolin clay. The first one is a natural clay, while the second one is a reconstituted clay. The objectives of such analyses were to quantify the sensitivity of each perturbed parameter to the calibrated reference parameters. The mean value of the error $(E_{mean})$ was calculated for each parameter as

$$E_{mean}(\%) = \frac{1}{N}\sum_{n=1}^{N}\left|\frac{q_{ref}^{k} - q_{per}^{k}}{q_{ref}^{k}}\right| \times 100 \tag{111}$$

where $N$ is the number of error calculations at different strain values; and $q_{ref}^{k}$ and $q_{per}^{k}$ are the predicted values of the deviatoric stress at $k_{th}$ strain using a reference parameter and a perturbed model parameter, respectively. The material parameters of the EVP models were divided into two groups: the Modified Cam Clay (MCC) parameters and the viscous parameters. Here, the parameters were perturbed by the same magnitude so that the influence of the perturbation of the parameter on model's prediction could be observed.

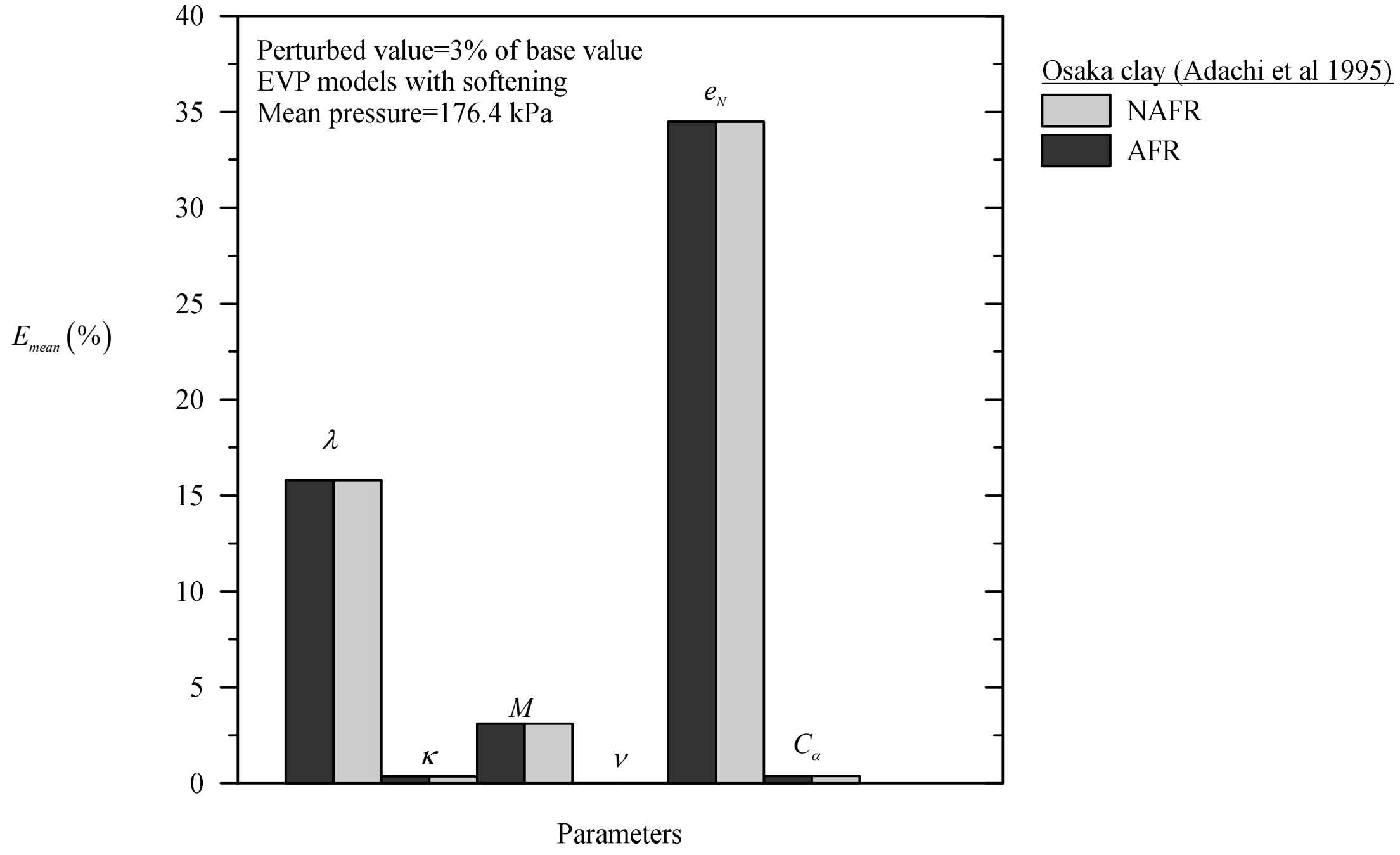

**Figure 9.** Sensitivity of perturbation for Osaka clay.

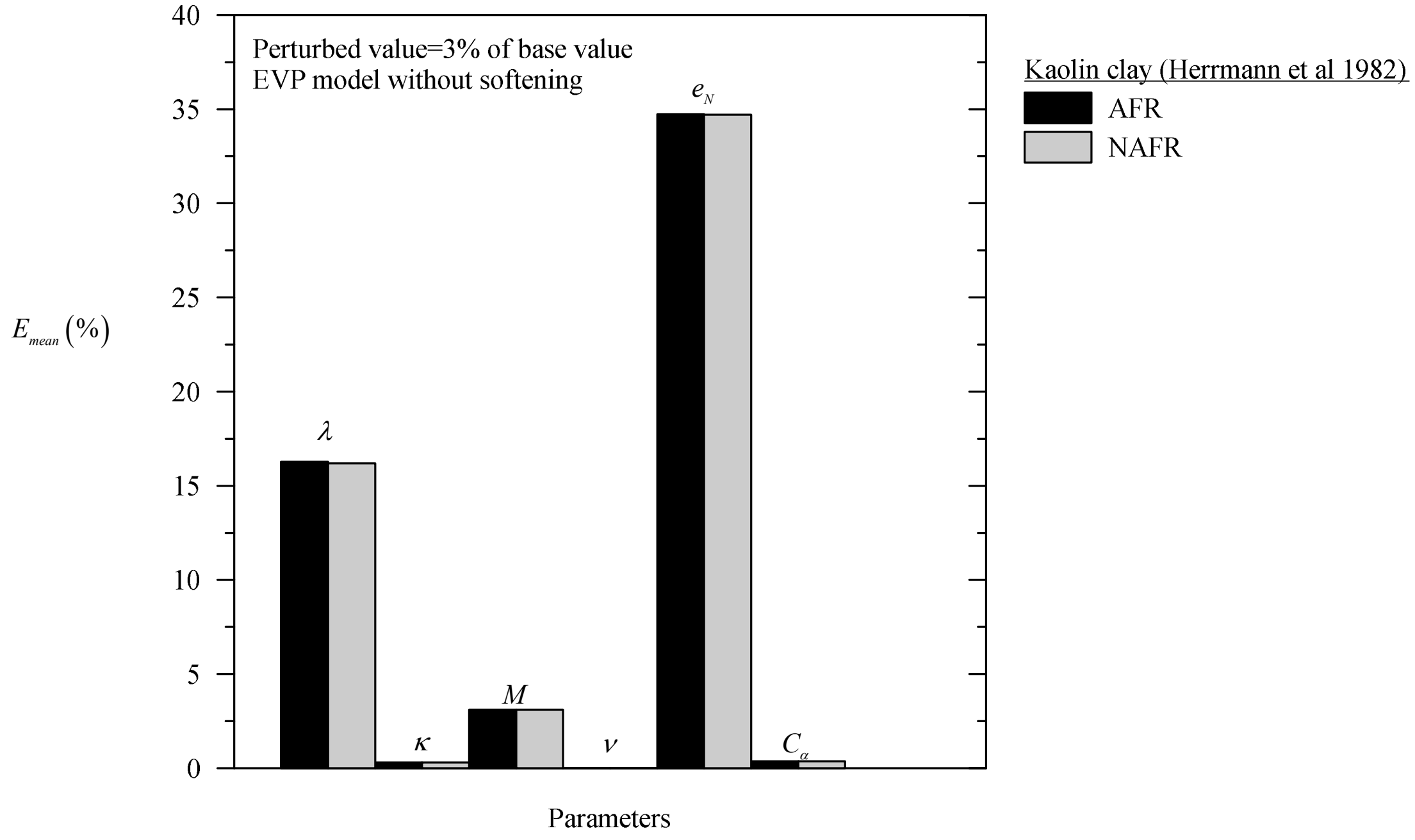

**Figure 10.** Sensitivity of perturbation for the Kaolin clay.

The parameters in the Osaka clay model were perturbed by 3% of their reference values, which were from experimental data. Then, for each parameter, $E_{mean}(\%)$ due to the perturbation was calculated for the associated flow rule and the non-associated flow rule EVP models, as shown in Equation (111). The perturbed results for the Osaka clay are shown in Figure 9. It was observed that parameters $e_N$ and $\lambda$ were most sensitive, while $M$ had medium sensitivity and $\kappa$ and $C_\alpha$ were less sensitive. To compare the perturbation influence on the flow rule, identical perturbation was also conducted for the Kaolin clay, as shown in Figure 10. In the perturbation analyses with the same magnitude, the effect of the flow rule was negligible.

## 7. Concluding Remarks

To describe the time-dependent behavior of clay, a new EVP model is formulated in this paper. The model requires six independent parameters, which are defined in the literature and can be extracted from the simple oedometer tests and the triaxial tests. A generalized nonlinear creep function is also used. The model captured a wide range of results when compared with experimental observations obtained from the drained and the undrained conditions for the triaxial compression test and the extension test, the creep test, the relaxation test, the over consolidation ratio effect test, and the strain rate test. Both undisturbed clay and reconstituted clay were considered in our study. From the comparison of the model predictions, it is evident that reasonably good agreement was obtained.

**Author Contributions:** Conceptualization, M.M.; Investigation, M.N.I.; Methodology, M.N.I. and M.M.; Supervision, C.T.G.; Validation, M.N.I.; Writing – original draft, M.N.I.; Writing – review & editing, M.M.

**Funding:** This research received no external funding.

**Acknowledgments:** The first author was financially supported during his research at the University of New South Wales, Canberra, Australia.

**Conflicts of Interest:** The authors declare no conflict of interest.

## Appendix A. Derivation of the Image Parameters

In this paper, we use the "radial mapping" rule [37] as discussed by Hashiguchi and Ueno [60] and Kaliakin and Dafalias [59] with the projection center of the yield locus at the center of the stress space (see Figure 1), which is similar to the Modified Cam Clay model (see also Desai and Siriwardane [9] (p.288, Figure 11.6); Islam and Gnanendran [22] (Appendix III); Kaliakin and Dafalias [59] (Equation 4b))

To obtain the image stress of the loading surface $(p, q)$ (see also Equation (41)) on the reference surface $(p_r, q_r)$ (see also Equation (40)), we use a parameter $(\beta_1)$ defined as

$$\left.\begin{aligned}(\sigma_r)_{ij} &= \beta_1\sigma_{ij}\\ p_r &= \beta_1 p\\ q_r &= \beta_1 q\end{aligned}\right\} \tag{A1}$$

Now, substituting Equation (A1) into Equation (40) and solving it for the quadratic real value, we obtain $\beta_1$ as

$$\beta_1 = \frac{p_{cr}}{p_{cl}} \tag{112}$$

Similarly, for the potential surface we obtain

$$\left.\begin{aligned} (\sigma_p)_{ij} &= \beta_2 \sigma_{ij} \\ p_p &= \beta_2 p \\ q_p &= \beta_2 q \end{aligned}\right\} \tag{113}$$

$$\beta_2 = \frac{p_{cp}}{p_{cl}} \tag{114}$$

where $p_{cl}, p_{cr}$ and $p_{cp}$ are defined in Equations (45), (46), and (47), respectively.

## Appendix B. Derivation of the Rate Sensitivity Function $\psi(F)$

Islam and Gnanendran [22] presented the derivation of the rate sensitivity function, $\psi(\boldsymbol{F})$ for the associated flow rule and also discussed the limitations of these approaches. For the sake of completeness, we provide a summary of the derivation for the non-associated flow rule. From Islam and Gnanendran [22], the viscoplastic strain rate $\left(\dot{\varepsilon}_v^{vp}\right)$ is given as

$$\dot{\varepsilon}_v^{vp} = \frac{\alpha}{\bar{t}(1+e_0)}\left(\frac{p_{cl}}{p_{cr}}\right)^{\frac{\lambda-\kappa}{\alpha}} \tag{115}$$

If we decompose $\dot{\varepsilon}_{ij}^{vp}$ into the volumetric $\left(\dot{\varepsilon}_v^{vp}\right)$ and the deviatoric $\left(\dot{\varepsilon}_q^{vp}\right)$ parts, the expression of the viscoplastic strain rate $\left(\dot{\varepsilon}_{ij}^{vp}\right)$ in the conventional triaxial stress space can be shown to be

$$\left.\begin{aligned} \dot{\varepsilon}_v^{vp} &= \psi \frac{\partial f_p}{\partial p_p} \\ \dot{\varepsilon}_q^{vp} &= \psi \frac{\partial f_p}{\partial q_p} \end{aligned}\right\} \tag{116}$$

If we compare Equations (A5) and (A6), the following expression can be obtained for the one-dimensional compression case considering the associated flow rule [22]:

$$\psi = \frac{\alpha_0}{\bar{t}(1+e_0)}\left(\frac{p_{cl}}{p_{cr}}\right)^{\frac{\lambda-\kappa}{\alpha}} \frac{1}{2p_{cp}\left[\frac{1}{\varsigma}-\frac{1}{2}\right]} \tag{117}$$

$$\varsigma = \frac{p_{cp}}{p_p} = 1 + \left(\frac{\eta_0}{M}\right)^2$$

$$\eta_0 = -\frac{6(\lambda-\kappa) - 2\sqrt{9(\lambda-\kappa)^2 + (2\lambda M)^2}}{4\lambda}$$

## Appendix C. Derivation of the Viscoplastic Strain Rate $\dot{\varepsilon}_{ij}^{vp}$

Recalling Equation (38), the viscoplastic strain rate $\left(\dot{\varepsilon}_{ij}^{vp}\right)$ can be written as

$$\dot{\varepsilon}_{ij}^{vp} = \langle\psi(F)\rangle \frac{\partial f_p}{\partial \sigma'_{ij}} \tag{118}$$

The expression for $\langle\psi(F)\rangle$ can be obtained from Appendix B. The mathematical form of $\frac{\partial f_p}{\partial \sigma'_{ij}}$ can be presented in terms of the chain rule:

$$\frac{\partial f_p}{\partial \sigma_{ij}^p} = \frac{\partial f_p}{\partial p_p}\frac{\partial p_p}{\partial \sigma_{ij}^p} + \frac{\partial f_p}{\partial q_p}\frac{\partial q_p}{\partial \sigma_{ij}^p} + \frac{\partial f_p}{\partial M}\frac{\partial M}{\partial b}\frac{\partial b}{\partial \sigma_{ij}^p} \quad (119)$$

In Equation (A9), the expression of the third term on the right-hand side depends on the definition of $M$, which is discussed in Section 3. The different terms in Equation (A9) can be expressed as

$$\frac{\partial p_p}{\partial \sigma_{ij}^p} = \frac{1}{3}\delta_{ij},\ where\ \delta_{ij} = \begin{cases} 1 & if \quad i = j \\ 0 & if \quad i \neq j \end{cases}$$

$$\frac{\partial q_p}{\partial \sigma_{ij}^p} = \begin{cases} \frac{3}{2q_p}\left(\sigma_{ij}^p - p_p\delta_{ij}\right) & if \quad i = j \\ \frac{3}{2q_p}\left(2\sigma_{ij}^p\right) & if \quad i \neq j \end{cases}$$

$$\frac{\partial f_p}{\partial M} = \frac{-2q_p^2}{M^3}$$

$$\frac{\partial M}{\partial b} = \frac{3sin\phi(2b-1)}{\left(\sqrt{b^2-b+1}\right)[3+(2b-1)sin\phi]} - \frac{12sin^2\phi\left(\sqrt{b^2-b+1}\right)}{[3+(2b-1)sin\phi]^2}$$

$$\frac{\partial b}{\partial \sigma_{11}^p} = -\frac{\sigma_{22}^p - \sigma_{33}^p}{\left(\sigma_{11}^p - \sigma_{33}^p\right)^2}$$

$$\frac{\partial b}{\partial \sigma_{22}^p} = \frac{1}{\sigma_{11}^p - \sigma_{33}^p}$$

$$\frac{\partial b}{\partial \sigma_{33}^p} = -\frac{1}{\sigma_{11}^p - \sigma_{33}^p} + \frac{\sigma_{22}^p - \sigma_{33}^p}{\left(\sigma_{11}^p - \sigma_{33}^p\right)^2}$$

In addition, following Lade [61], the third term in Equation (A9), also can be expressed in terms of Lode angle ($\theta$) as follows

$$\frac{\partial M}{\partial \theta} = -\frac{1}{2}\frac{k\ sin(3\theta)cos\left(\frac{\pi}{6} + \frac{1}{3}cos^{-1}\left(cos(3\theta)\right)\sqrt{\frac{k}{27}}\right)}{sin\left(\frac{\pi}{6} + \frac{1}{3}cos^{-1}\left(cos(3\theta)\right)\sqrt{\frac{k}{27}}\right)^2\sqrt{-3cos(3\theta)^2k + 81}}$$

$$\frac{\partial \theta}{\partial \sigma_1^p} = \frac{1}{3}cos^{-1}\left[\frac{\frac{1}{2}\frac{\frac{1}{\sigma_1^p - \sigma_3^p}(-6b^3+6b^2+3b)}{(b^2-b+1)^{\frac{3}{2}}} - \frac{3}{4}\frac{1}{(b^2-b+1)^{\frac{5}{2}}}}{\left\{(2b^3-3b^2-3b+2)\left(\frac{1}{\sigma_1^p - \sigma_3^p}(-2b^2+b)\right)\right\}}\right]$$

$$\frac{\partial \theta}{\partial \sigma_2^p} = \frac{1}{3}cos^{-1}\left[\frac{\frac{1}{2}\frac{\frac{1}{\sigma_1^p - \sigma_3^p}(6b^2-6b-3)}{(b^2-b+1)^{\frac{3}{2}}} - \frac{3}{4}\frac{1}{(b^2-b+1)^{\frac{5}{2}}}}{\left\{(2b^3-3b^2-3b+2)\left(\frac{1}{\sigma_1^p - \sigma_3^p}(2b-1)\right)\right\}}\right]$$

$$\frac{\partial\theta}{\partial\sigma_3^p} = \frac{1}{3}cos^{-1}\left[\begin{array}{c}\dfrac{1}{2}\dfrac{\dfrac{1}{\sigma_1^p-\sigma_3^p}(6b^3-12b^2+3b+3)}{(b^2-b+1)^{\frac{3}{2}}} - \dfrac{3}{4}\dfrac{1}{(b^2-b+1)^{\frac{5}{2}}} \\ \left\{(2b^3-3b^2-3b+2)\left(\dfrac{1}{\sigma_1^p-\sigma_3^p}(2b^2-3b+1)\right)\right\}\end{array}\right]$$

## Appendix D. Derivation of the Gradient Matrix $\boldsymbol{H_n}$

Owen and Hinton [8] presented the definition of the gradient matrix as

$$\boldsymbol{H}_n = \left(\frac{\partial\dot{\boldsymbol{\varepsilon}}^{vp}}{\partial\boldsymbol{\sigma}'}\right)_n \tag{A10}$$

where $n$ is the number of dimensions. For two-dimensional cases, the total number of elements are sixteen, which can be expressed as

$$[\boldsymbol{H}] = \begin{bmatrix} \frac{\partial\dot{\varepsilon}_{11}^{vp}}{\partial\sigma_{11}} & \frac{\partial\dot{\varepsilon}_{11}^{vp}}{\partial\sigma_{22}} & \frac{\partial\dot{\varepsilon}_{11}^{vp}}{\partial\sigma_{33}} & \frac{\partial\dot{\varepsilon}_{11}^{vp}}{\partial\sigma_{12}} \\ \frac{\partial\dot{\varepsilon}_{22}^{vp}}{\partial\sigma_{11}} & \frac{\partial\dot{\varepsilon}_{22}^{vp}}{\partial\sigma_{22}} & \frac{\partial\dot{\varepsilon}_{22}^{vp}}{\partial\sigma_{33}} & \frac{\partial\dot{\varepsilon}_{22}^{vp}}{\partial\sigma_{12}} \\ \frac{\partial\dot{\varepsilon}_{33}^{vp}}{\partial\sigma_{11}} & \frac{\partial\dot{\varepsilon}_{33}^{vp}}{\partial\sigma_{22}} & \frac{\partial\dot{\varepsilon}_{33}^{vp}}{\partial\sigma_{33}} & \frac{\partial\dot{\varepsilon}_{33}^{vp}}{\partial\sigma_{12}} \\ \frac{\partial\dot{\varepsilon}_{12}^{vp}}{\partial\sigma_{11}} & \frac{\partial\dot{\varepsilon}_{12}^{vp}}{\partial\sigma_{22}} & \frac{\partial\dot{\varepsilon}_{12}^{vp}}{\partial\sigma_{33}} & \frac{\partial\dot{\varepsilon}_{12}^{vp}}{\partial\sigma_{12}} \end{bmatrix} \tag{120}$$

Applying the chain rule to the derivatives $\left(\frac{\partial\dot{\varepsilon}_{ij}^{vp}}{\partial\sigma_{ij}}\right)$ of the strain rate with respect to the stress components, we have

$$\left(\frac{\partial\dot{\varepsilon}_{ij}^{vp}}{\partial\sigma_{ij}} = \frac{\partial\dot{\varepsilon}_{ij}^{vp}}{\partial p}\frac{\partial p}{\partial\sigma_{ij}} + \frac{\partial\dot{\varepsilon}_{ij}^{vp}}{\partial q}\frac{\partial q}{\partial\sigma_{ij}}\right) \tag{A121}$$

Implementing Equation (A12) into Equation (A11), we obtain the complete form of the gradient matrix for two-dimensional case:

$$[\boldsymbol{H}] = \begin{bmatrix} \frac{\partial\dot{\varepsilon}_{11}^{vp}}{\partial p}\cdot\frac{\partial p}{\partial\sigma_{11}} + \frac{\partial\dot{\varepsilon}_{11}^{vp}}{\partial q}\cdot\frac{\partial q}{\partial\sigma_{11}} & \frac{\partial\dot{\varepsilon}_{11}^{vp}}{\partial p}\cdot\frac{\partial p}{\partial\sigma_{22}} + \frac{\partial\dot{\varepsilon}_{11}^{vp}}{\partial q}\cdot\frac{\partial q}{\partial\sigma_{22}} & \frac{\partial\dot{\varepsilon}_{11}^{vp}}{\partial p}\cdot\frac{\partial p}{\partial\sigma_{33}} + \frac{\partial\dot{\varepsilon}_{11}^{vp}}{\partial q}\cdot\frac{\partial q}{\partial\sigma_{33}} & \frac{\partial\dot{\varepsilon}_{11}^{vp}}{\partial p}\cdot\frac{\partial p}{\partial\sigma_{12}} + \frac{\partial\dot{\varepsilon}_{11}^{vp}}{\partial q}\cdot\frac{\partial q}{\partial\sigma_{12}} \\ \frac{\partial\dot{\varepsilon}_{22}^{vp}}{\partial p}\cdot\frac{\partial p}{\partial\sigma_{11}} + \frac{\partial\dot{\varepsilon}_{22}^{vp}}{\partial q}\cdot\frac{\partial q}{\partial\sigma_{11}} & \frac{\partial\dot{\varepsilon}_{22}^{vp}}{\partial p}\cdot\frac{\partial p}{\partial\sigma_{22}} + \frac{\partial\dot{\varepsilon}_{22}^{vp}}{\partial q}\cdot\frac{\partial q}{\partial\sigma_{22}} & \frac{\partial\dot{\varepsilon}_{22}^{vp}}{\partial p}\cdot\frac{\partial p}{\partial\sigma_{33}} + \frac{\partial\dot{\varepsilon}_{22}^{vp}}{\partial q}\cdot\frac{\partial q}{\partial\sigma_{33}} & \frac{\partial\dot{\varepsilon}_{22}^{vp}}{\partial p}\cdot\frac{\partial p}{\partial\sigma_{12}} + \frac{\partial\dot{\varepsilon}_{22}^{vp}}{\partial q}\cdot\frac{\partial q}{\partial\sigma_{12}} \\ \frac{\partial\dot{\varepsilon}_{33}^{vp}}{\partial p}\cdot\frac{\partial p}{\partial\sigma_{11}} + \frac{\partial\dot{\varepsilon}_{33}^{vp}}{\partial q}\cdot\frac{\partial q}{\partial\sigma_{11}} & \frac{\partial\dot{\varepsilon}_{33}^{vp}}{\partial p}\cdot\frac{\partial p}{\partial\sigma_{22}} + \frac{\partial\dot{\varepsilon}_{33}^{vp}}{\partial q}\cdot\frac{\partial q}{\partial\sigma_{22}} & \frac{\partial\dot{\varepsilon}_{33}^{vp}}{\partial p}\cdot\frac{\partial p}{\partial\sigma_{33}} + \frac{\partial\dot{\varepsilon}_{33}^{vp}}{\partial q}\cdot\frac{\partial q}{\partial\sigma_{33}} & \frac{\partial\dot{\varepsilon}_{33}^{vp}}{\partial p}\cdot\frac{\partial p}{\partial\sigma_{12}} + \frac{\partial\dot{\varepsilon}_{33}^{vp}}{\partial q}\cdot\frac{\partial q}{\partial\sigma_{12}} \\ \frac{\partial\dot{\varepsilon}_{12}^{vp}}{\partial p}\cdot\frac{\partial p}{\partial\sigma_{11}} + \frac{\partial\dot{\varepsilon}_{12}^{vp}}{\partial q}\cdot\frac{\partial q}{\partial\sigma_{11}} & \frac{\partial\dot{\varepsilon}_{12}^{vp}}{\partial p}\cdot\frac{\partial p}{\partial\sigma_{22}} + \frac{\partial\dot{\varepsilon}_{12}^{vp}}{\partial q}\cdot\frac{\partial q}{\partial\sigma_{22}} & \frac{\partial\dot{\varepsilon}_{12}^{vp}}{\partial p}\cdot\frac{\partial p}{\partial\sigma_{33}} + \frac{\partial\dot{\varepsilon}_{12}^{vp}}{\partial q}\cdot\frac{\partial q}{\partial\sigma_{33}} & \frac{\partial\dot{\varepsilon}_{12}^{vp}}{\partial p}\cdot\frac{\partial p}{\partial\sigma_{12}} + \frac{\partial\dot{\varepsilon}_{12}^{vp}}{\partial q}\cdot\frac{\partial q}{\partial\sigma_{12}} \end{bmatrix} \tag{A1223}$$

## References

1. Saito, M.; Uezawa, H. Failure of soil due to creep. In Proceedings of the 5th International Conference on Soil Mechanics and Foundation Engineering (ICSMFE); Paris, France, 17–22 July 1961; Volume 1, pp. 315–318.
2. Muller-Salzburg, L. *The Rock Slide in the Vaiont Valley*; Springer: Wien, Germany, 1964.
3. Rowe, P.W. The stress-dilatancy relation for static equilibrium of an assembly of particles in contact. *Proc. R. Soc. Lond. Ser. A Math. Phys. Sci.* **1962**, *269*, 500–527, doi:10.1098/rspa.1962.0193.
4. Terzaghi, K. *Theoretical Soil Mechanics*; Wiley: New York, NY, USA, 1943.

5. Barden, L. Consolidation of Clay with Non-linear Viscosity. *Géotechnique* **1965**, *15*, 345–362, doi:10.1680/geot.1965.15.4.345.
6. Alonso, E.E.; Gens, A.; Lloret, A. Precompression design for secondary settlement reduction. *Geotechnique* **2000**, *50*, 645–656.
7. Potts, D.M.; Zdravkovic, L. *Finite Element Analysis in Geotechnical Engineering: Theory*; Thomas Telford: London, UK, 1999.
8. Owen, D.R.J.; Hinton, E. *Finite Elements in Plasticity: Theory and Practice*; Pineridge Press Limited: Swansea, UK, 1980.
9. Desai, C.; Siriwardane, H.J. *Constitutive Laws For Engineering Materials*; Prentice-Hall: Upper Saddle River, NJ, USA, 1984.
10. Chaboche, J.L. A review of some plasticity and viscoplasticity constitutive theories. *Int. J. Plast.* **2008**, *24*, 1642–1693, doi:10.1016/j.ijplas.2008.03.009.
11. Tatsuoka, F.; Bendetto, H.D.; Enomoto, T.; Kawabe, S.; Kongkitkul, W. Various viscosity types of geomaterials in shear and their mathematical expression. *Soils Found.* **2008**, *48*, 41–60.
12. Liingaard, M.; Augustesen, A.; Lade, P. Characterization of Models for Time-Dependent Behavior of Soils. *Int. J. Geomech.* **2004**, *4*, 157–177, doi:10.1061/(ASCE)1532-3641(2004)4:3(157).
13. Singh, A.; Mitchell, J.K. General Stress-strain-time Function for Soils. *J. Soil Mech. Found. Div ASCE* **1968**, *94*, 21–46.
14. Feda, J. *Creep of Soils: And Related Phenomena (Developments in Geotechnical Engineering)*, 2nd ed.; Elsevier Science: Amsterdam, The Netherlands, 1992.
15. Adachi, T.; Okano, M. A constitutive equation for normally consolidated clay. *Soils Found.* **1974**, *14*, 55–73.
16. Sekiguchi, H. Macrometric Approaches-Static-Intricsically Time Dependent Constitutive Laws of Soils. In Proceedings of the 11th ICFMFE, San Francisco, CA, USA, 29–31 March 2019; pp. 66–98.
17. Sekiguchi, H. Rheological Characteristics of Clays. In Proceedings of the 9th International Conference on Soil Mechanics and Foundation Engineering, Tokyo, Japan, 10–15 July 1977; pp. 289–292.
18. Dafalias, Y.F.; Herrmann, L.R. Bounding surface formulation of soil plasticity. In *Soil Mechanics-Transient and Cyclic Loads*; Pande, G.N., Zienkiewicz, O.C., Eds.; Wiley: Chichester, UK, 1982; pp. 253–282.
19. Borja, R.I.; Kavazanjian, J.E. A constitutive model for the stress-strain&-time behaviour of 'wet' clays. *Geotechnique* **1985**, *35*, 283–298.
20. Perzyna, P. Constitutive equations for rate-sensitive plastic materials. *Q. Appl. Math.* **1963**, *20*, 321–331.
21. Roscoe, K.H.; Burland, J.B. On the generalized stress-strain behavior of wet clay. In *Engineering Plasticity*; Heyman, J., Leckie, F.A., Eds.; Cambridge University Press: Cambridge, UK, 1968; pp. 535–609.
22. Islam, M.N.; Gnanendran, C.T. Elastic-Viscoplastic Model for Clays: Development, Validation, and Application. *J. Eng. Mech.* **2017**, *143*, doi:10.1061/(asce)em.1943–7889.0001345.
23. Carter, J.P.; Balaam, N.P. *AFENA User's Manual. Version 5.0*; Center for Geotechnical Research, University of Sydney: Sydney, Australia, 1995.
24. Atkin, R.J.; Craine, R.E. Continuum theories of mixtures: Basic theory and historical development. *Q. J. Mech. Appl. Math.* **1976**, *29*, 209–244.
25. Bowen, R.M. Theory of Mixtures, Part I. In *Continuum Physics III*; Eringen, A., Ed., Academic Press: Cambridge, MA, USA, 1976, pp. 2–127.
26. Truesdell, C. *Rational Thermodynamics*, 2nd ed.; Springer: New York, NY, USA, 1984.
27. Rajagopal, K.R. On a hierarchy of approximate models for flows of incompressible fluids through porous solids. *Math. Models Methods Appl. Sci.* **2007**, *17*, 215–252.
28. Martins-Costa, M.L.; Sampaio, R.; da Gama, R.M.S. Modelling and simulation of energy transfer in a saturated flow through a porous medium. *Appl. Math. Model.* **1992**, *16*, 589–597, doi:10.1016/0307-904X(92)90034-Z.
29. Oka, F.; Kimoto, S. *Computational Modeling of Multiphase Geomaterials*; CRC Press: London, UK, 2013.
30. Massoudi, M. On the importance of material frame-indifference and lift forces in multiphase flows. *Chem. Eng. Sci.* **2002**, *57*, 3687–3701, doi:10.1016/S0009-2509(02)00237-3.
31. Massoudi, M. Constitutive relations for the interaction force in multicomponent particulate flows. *Int. J. Non-Linear Mech.* **2003**, *38*, 313–336, doi:10.1016/S0020-7462(01)00064-6.
32. Williams, W.O. Constitutive equations for flow of an incompressible viscous fluid through a porous medium. *Q. Appl. Math.* **1978**, *36*, 255–267.

33. Lewis, R.W.; Schrefler, B.A. *The Finite Element Method in the Static and Dynamic Deformation and Consolidation of Porous Media*, 2nd ed.; John Wiley & Sons: New York, NY, USA, 1998.
34. Davis, R.O.; Selvadurai, A.P.S. *Plasticity and Geomechanics*; Cambridge University Press: New York, N, USA, 2002.
35. Schofield, A.N.; Wroth, P. *Critical State Soil Mechanics*; McGrawHill: London, UK, 1968.
36. Matsuoka, H.; Sun, D.A. *The SMP Concept-Based 3D Constitutive Models for Geomaterials*; Taylor & Francis: New York, NY. USA, 2006.
37. Phillips, A.; Sierakowski, R.L. On the concept of the yield surface. *Acta Mech.* **1965**, *1*, 29–35, doi:10.1007/bf01270502.
38. Habib, P. Influence of the variation of the average principal stress upon the shearing strength of soils. In Proceedings of the 3rd International Conference on Soil Mechanics and Foundation Engineering, Zurich, Switzerland, 16–27 August 1953; pp. 131–136.
39. Bjerrum, L. Engineering geology of Norwegian normally consolidated marine clays as related to settlements of buildings. *Geotechnique* **1967**, *17*, 81–118.
40. Zienkiewicz, O.C.; Humpheson, C.; Lewis, R.W. Associated and non-associated visco-plasticity and plasticity in soil mechanics. *Geotechnique* **1975**, *25*, 671–689.
41. Maranini, E.; Yamaguchi, T. A non-associated viscoplastic model for the behaviour of granite in triaxial compression. *Mech. Mater.* **2001**, *33*, 283–293, doi:10.1016/S0167-6636(01)00052-7.
42. Kutter, B.L.; Sathialingam, N. Elastic-viscoplastic modelling of the rate-dependent behaviour of clays. *Géotechnique* **1992**, *42*, 427–441.
43. Hickman, R.J.; Gutierrez, M.S. Formulation of a three-dimensional rate-dependent constitutive model for chalk and porous rocks. *Int. J. Numer. Anal. Methods Geomech.* **2007**, *31*, 583–605, doi:10.1002/nag.546.
44. Cristescu, N.D. Nonassociated elastic/viscoplastic constitutive equations for sand. *Int. J. Plast.* **1991**, *7*, 41–64, doi:10.1016/0749-6419(91)90004-I.
45. Florea, D. Nonassociated elastic/viscoplastic model for bituminous concrete. *Int. J. Eng. Sci.* **1994**, *32*, 87–93, doi:10.1016/0020-7225(94)90151-1.
46. Jin, J.; Cristescu, N.D. An elastic/viscoplastic model for transient creep of rock salt. *Int. J. Plast.* **1998**, *14*, 85–107.
47. Augustesen, A.; Liingaard, M.; Lade, P. Evaluation of Time-Dependent Behavior of Soils. *Int. J. Geomech.* **2004**, *4*, 137–156, doi:10.1061/(ASCE)1532-3641(2004)4:3(137).
48. Bear, J.; Bachmat, Y. *Introduction to Modeling of Transport Phenomena in Porous Media*; Kluwer Academic Publishers: London, UK, 1990; Volume 4.
49. Zienkiewicz, O.C.; Taylor, R.L.; Zhu, J.Z. *The Finite Element Method: Its Basis and Fundamentals*, 6th ed.; Elsevier: New York, NY, USA, 2005.
50. Galerkin, B.G. Series solution of some problems of elastic equilibrium of rods and plates (in Russian). *Vestn. Inzh. Tech.* **1915**, *19*, 897–908.
51. Koutromanos, I. *Fundamentals of Finite Element Analysis: Linear Finite Element Analysis*; John Wiley and Sons: Noida, India, 2018.
52. Kanchi, M.B.; Zienkiewicz, O.C.; Owen, D.R.J. The visco-plastic approach to problems of plasticity and creep involving geometric non-linear effects. *Int. J. Numer. Methods Eng.* **1978**, *12*, 169–181.
53. Adachi, T.; Oka, F.; Hirata, T.; Hashimoto, T.; Nagaya, J.; Mimura, M.; Pradhan, T.B.S. Stress–strain behavior and yielding characteristics of Eastern Osaka clay. *Soil Found.* **1995**, *35*, 1–13.
54. Lacerda, W.A. Stress-Relaxation and Creep Effects on Soil Deformation. Ph.D. Thesis, University of California, Berkeley, CA, USA, 1976.
55. Herrmann, L.R.; Shen, C.K.; Jafroudi, S.; DeNatale, J.S.; Dafalias, Y.F. *A Verification Study for the Bounding Surface Plasticity Model for Cohesive Soils*; Final report to the Civil Engineering Laboratory; Naval Construction Battalion Center: Port Hueneme, CA, USA, 1982.
56. Yin, J.H.; Zhu, J.G. Measured and predicted time-dependent stress-strain behaviour of Hong Kong marine deposits. *Can. Geotech. J.* **1999**, *36*, 760–766, doi:10.1139/t99-043.
57. Jiang, J.; Ling, H.I.; Kaliakin, V.N. An Associative and Non-Associative Anisotropic Bounding Surface Model for Clay. *J. Appl. Mech.* **2012**, *79*, 031010, doi:10.1115/1.4005958.
58. Whittle, A.; Kavvadas, M. Formulation of MIT-E3 Constitutive Model for Overconsolidated Clays. *J. Geotech. Eng.* **1994**, *120*, 173–198.

59. Kaliakin, V.N.; Dafalias, Y.F. Verification of the Elastoplastic-Viscopalstic Bounding Surface Model for Cohesive Soils. *Soils Found.* **1990**, *30*, 25–36.
60. Hashiguchi, K.; Ueno, M. Elastoplastic constitutive laws of granular materials. In Proceedings of the 9th International Conference on Soil Mechanics and Foundation Engineering (ICFSME), Session 9, Tokyo, Japan, 10–15 July 1977, pp. 73–82.
61. Lade, P.V. *Triaxial Testing of Soils*; Chapter 2; John Wiley & Sons: Chichester, UK, 2016; p. 88.